\documentstyle[12pt,graphicx,epsf]{article}
\begin{document}

\def\NPB#1#2#3{{\it Nucl.\ Phys.}\/ {\bf B#1} (#2) #3}
\def\PLB#1#2#3{{\it Phys.\ Lett.}\/ {\bf B#1} (#2) #3}
\def\PRD#1#2#3{{\it Phys.\ Rev.}\/ {\bf D#1} (#2) #3}
\def\PRL#1#2#3{{\it Phys.\ Rev.\ Lett.}\/ {\bf #1} (#2) #3}
\def\PRT#1#2#3{{\it Phys.\ Rep.}\/ {\bf#1} (#2) #3}
\def\MODA#1#2#3{{\it Mod.\ Phys.\ Lett.}\/ {\bf A#1} (#2) #3}
\def\IJMP#1#2#3{{\it Int.\ J.\ Mod.\ Phys.}\/ {\bf A#1} (#2) #3}
\def\nuvc#1#2#3{{\it Nuovo Cimento}\/ {\bf #1A} (#2) #3}
\def\RPP#1#2#3{{\it Rept.\ Prog.\ Phys.}\/ {\bf #1} (#2) #3}
\def\APJ#1#2#3{{\it Astrophys.\ J.}\/ {\bf #1} (#2) #3}
\def\APP#1#2#3{{\it Astropart.\ Phys.}\/ {\bf #1} (#2) #3}
\def\etal{{\it et al\/}}

\newcommand{\bev}{\begin{verbatim}}
\newcommand{\beq}{\begin{equation}}
\newcommand{\beqa}{\begin{eqnarray}}
\newcommand{\beqn}{\begin{eqnarray}}
\newcommand{\eeqn}{\end{eqnarray}}
\newcommand{\eeqa}{\end{eqnarray}}
\newcommand{\eeq}{\end{equation}}
\newcommand{\Eev}{\end{verbatim}}
\newcommand{\bec}{\begin{center}}
\newcommand{\eec}{\end{center}}
\def\ie{{\it i.e.}}
\def\eg{{\it e.g.}}
\def\half{{\textstyle{1\over 2}}}
\def\nicefrac#1#2{\hbox{${#1\over #2}$}}
\def\third{{\textstyle {1\over3}}}
\def\quarter{{\textstyle {1\over4}}}
\def\m{{\tt -}}
\def\mass{M_{l^+ l^-}}
\def\p{{\tt +}}

\def\slash#1{#1\hskip-6pt/\hskip6pt}
\def\slk{\slash{k}}
\def\GeV{\,{\rm GeV}}
\def\TeV{\,{\rm TeV}}
\def\y{\,{\rm y}}

\def\l{\langle}
\def\r{\rangle}

\begin{titlepage}
\samepage{
\setcounter{page}{1}
\rightline{UNILE-CBR-2001-3}
\rightline{OUTP-01-33P}
\vspace{1.5cm}
\begin{center}
 {\Large \bf SUSY QCD and High Energy Cosmic Rays 1.\\}
\vspace{.5 cm}
{\Large \bf  Fragmentation functions of SUSY QCD \\  }
\vspace{1.cm}
 {\large Claudio Corian\`{o}\footnote{ E-mail address:
Claudio.Coriano@le.infn.it}
 and Alon E. Faraggi\footnote{faraggi@thphys.ox.ac.uk}\\
\vspace{.5cm}
{\it $^1$Dipartimento di Fisica\\
 Universita' di Lecce\\
 I.N.F.N. Sezione di Lecce \\
Via Arnesano, 73100 Lecce, Italy\\}
\vspace{.5cm}
{\it $^2$Theoretical Physics Department\\
University of Oxford, Oxford, OX1 3NP, United Kingdom}}
\end{center}
\begin{abstract}
The supersymmetric evolution of the fragmentation functions (or timelike
evolution)
within $N=1$ $QCD$ is discussed and predictions for the fragmentation functions
of the theory (into final protons) are given.
We use a backward running of the supersymmetric DGLAP equations,
using a method developed in previous works.
We start from the usual QCD parameterizations at low energy and run the DGLAP
back,
up to an intermediate scale - assumed to be supersymmetric - where we switch-on
supersymmetry.
{}From there on we assume the applicability of an $N=1$ supersymmetric
evolution (ESAP).
We elaborate on possible application of these results to High Energy
Cosmic Rays near the GZK cutoff.

\end{abstract}
\smallskip}
\end{titlepage}

\section{Introduction}

The Standard Model of elementary particle physics successfully accounts for
all the observed particle properties. Its remarkable success
suggests the possibility that it correctly describes the particle
properties up to energy scales which far exceed the currently accessible
energy scales.
Indeed, this possibility has been entertained in the explorations of
Grand Unified, and Superstring, Theories. Additionally, the spectacular
confirmation of the Standard Model in high energy colliders, as well
as the proton longevity and suppression of left--handed neutrino masses,
provide strong support for the grand desert scenario and unification.
The question then is how are we going to find out whether this is
indeed the path chosen by nature, as direct signatures at LHC/NLC
will be able to probe the desert only up to a few TeV.
Therefore we need new tools to achieve this.
An example of such a probe that has been used with great success is that
of the gauge coupling unification. The general methodology in this respect
is to extract the gauge couplings from experiments that are performed
at the accessible energy scales. Based on concrete assumptions in
regard to the particle content in the desert, the couplings are then
extrapolated to the high energy scale and the unification hypothesis is
tested. In this manner the consistency of the specific assumptions in
regard to the particle content in the desert with the hypothesis of the
gauge coupling unification is subjected to an experimental test.
The well known spectacular success of this methodology is in
differentiating between gauge coupling unification in supersymmetric
and non--supersymmetric Grand Unified Theories
\cite{gcurge}. The unification hypothesis is
found not to be consistent with the low energy data, unless the
minimal particle content is supersymmetric, or is modified
in some other way. Our task then is to develop additional
tools that can serve as similar useful probes of the desert.

\subsection{ Motivations for this work}
In this paper we start applying this philosophy to another physical setting.

A different experimental observation - which also points toward
the validity of the big desert scenario -
is the existence of Ultra High Energy
Cosmic Rays above the Greisen--Zatsepin--Kuzmin cutoff.
A plausible explanation for the observation of such cosmic rays
is the existence of supermassive metastable states with mass
of the order $10^{12}-10^{13}{\rm GeV}$, and a lifetime that exceeds
$10^{10}y$. The possibility that
such states compose a substantial component of the dark matter
as well as that they may explain the observation of UHECR
has been discussed elsewhere. It has been further shown
that realistic string models often give rise to states
with precisely such properties \cite{ccf}.

The existence of such super--massive, slowly decaying states,
may therefore serve as a probe of the grand desert.
A plausible explanation of the observed showers in the UHECR
is that the original supermassive decaying particle decays
into strongly interacting particles which subsequently fragment
into hadrons \cite{Subir1}. This hypothesis therefore needs
the relevant fragmentation functions at the energy scale
of the originally decaying particle. These functions,
however, similarly to the case of the gauge couplings
considered in much of the
literature on unification, are extracted from
experiments at the currently accessible (low) energy scales which are much
below that of the decaying particle.
It is obvious that this points toward the use of the
renormalization group (RG)
evolution.
However, this
evolution, as well as the composition of the produced shower,
depends on the assumptions made in regard to the composition of
the spectrum in the desert. We conclude that similarly to the gauge coupling
unification methodology, the evolution of the fragmentation functions
by utilizing the renormalization group extrapolation, and the
subsequent implementation in the analysis of the UHECR hadronic
showers, can serve as a tool that differentiates between different
assumptions on the very high energy spectrum.

With these motivations in mind, here we develop the instruments to evolve the
fragmentation functions
in supersymmetric QCD up to a high energy scale.
In particular, in this paper we present
for the first time the structure of the supersymmetric DGLAP
(Exact Supersymmetric DGLAP or ESAP)
equations in the timelike region.
We solve them numerically for all flavours
and introduce various combinations of non-singlet matrix equations
to reach this result. We then study the impact of the
new evolution assuming a common scale for SUSY breaking and restoration,
parameterized by the mass of the superpartners, here assumed to be mass
degenerate.

\subsection{Step Approximations}
Given the long stretch it takes to proceed with the analysis
of the SUSY evolution, a topic of numerical complexity in its own,
the applications to cosmic rays of our work will be presented in a companion
paper that will follow shortly. We also have decided to address issues related
to
the fragmentation region of $N=1$ QCD starting from the ``low energy'' end,
since nothing is known of these functions at any scale.
The term ``low energy''
(the results we discuss here are obtained in the $10^{3}- 10^{8}$ GeV energy
range )
- from the point of view of collider phenomenology - is a misnomer,
but it is not in the context of high energy cosmic ray physics.
There is also another more direct reason for limiting our analysis to this
range.
We have discovered some features of the RG evolution which require
a special care at such large energies. We find -in our numerical studies-
the onset of an instability in the evolution in the singlet sector of the
fragmentation which requires an independent investigation. It is not clear to
us, at this point,
whether this instability is an intrinsic limitation of the algorithm used
in the numerical implementation or if the perturbative
expansion needs a resummation. At such large scales this last option remains
open.
We should also mention that we work in ``x'' space, and this might be a
limitation.
We hope to return to this issue in the near future.

 The analysis of fragmentation functions of SUSY QCD is a topic
of remarkable phenomenological interest both
for collider phenomenology and in the astroparticle context.
While - recently - a detailed analysis of the
supersymmetric evolution of the parton distributions of SUSY QCD has been
presented
\cite{CC1,CC2}, the study of the evolution of the supersymmetric
fragmentation functions is still missing. Of particular interest
is the study of a combined QCD/SQCD evolution with intermediate
regions in the evolution
characterized by partial supersymmetry (SAP) or exact supersymmetry (ESAP)
\cite{CC2}.
It is well known that a matching
between these regions is possible with specific boundary conditions. These
boundary conditions should come from some
deeper understanding of the way in which supersymmetry is broken and restored
as we run into the different stages of the renormalization group evolution.
In a first approximation, however, it is possible to assume
that the regular QCD distributions (or fragmentation) functions are continuous
at each intermediate region,
thereby neglecting threshold effects which cannot be inferred
from first principles. As we cross any supersymmetric region, starting,
for instance, from the end of the regular QCD evolution,
it is possible to assume that supersymmetric distributions and fragmentation
functions are generated radiatively. We recall that a similar modeling
-which neglects possible effects of higher twist at the thresholds- is well
spread in the
case of ordinary QCD.
\section{Initial/Final State Scaling Violations}
\subsection{The Spacelike Evolution}
The first analysis of the spacelike evolution was
carried out in ref.\cite{KR},
where simple models -obtained from the analysis of the first 2 moments -
of the supersymmetric parton distributions were also presented.
A complete strategy to solve these equations and the strategy to generate
supersymmetric scaling violations in N=1 QCD was put forward in
Refs.~\cite{CC1,CC2}.

In $N=1$ QCD gluons have partners called gluinos (here denoted by $\lambda$)
and left- and right-handed quarks have complex scalar partners (squarks) which
we denote as
$ \tilde{q_L}$ and $\tilde{q_R}$ with $\tilde{q}=\tilde{q}_L + \tilde{q}_R$
(for left-handed and right-handed squarks respectively).

The interaction between the elementary fields are described by the
$SU(3)$ color gauge invariant and supersymmetric Lagrangian

\beqa
&&{ \cal L}= -{1\over 4} G_{\mu\nu}^a G^{\mu\nu}_a +
{1\over 2}\bar{\lambda}_a \left( i \not{D}\right) \lambda_a \nonumber \\
&& + i \bar{q}_i \not{D} q_i + D_\mu \tilde{q}_R D^\mu\tilde{q}^{R}
+ D_\mu \tilde{q}_L D^\mu\tilde{q}_{L} + i g \sqrt{2}
\left(\bar{\lambda^a}_R \tilde{q}_{i L}^\dagger  T^a q_{L i} +
\bar{\lambda^a}_L \tilde{q}_{i R}^\dagger  T^a q_{R i}  - {\tt h.\,\,\,c.}
\right)
\nonumber \\
&& -{1\over 2} g^2 \left( \tilde{q}_{L i}^\dagger T^a \tilde{q}_{L i}
- \tilde{q}_{R i}^\dagger T^a \tilde{q}_{R i}^\dagger\right)^2
+ {\tt mass\,\,\,\, terms },\nonumber \\
\eeqa

where $a$ runs over the adjoint of the color group and $i$ denotes the
flavours.

$\tilde{q}_{iR}$ are the supersymmetric partners of right-handed quarks
$q_{iR}$ and
$\tilde{q}_{iL}$ are those of the left-handed quarks $q_{iL}$.

We remind here that in actual phenomenological applications,
it is useful to extract the light-cone dynamics -or parton model picture-
of supersymmetric collisions by invoking a factorization of the cross section
into hard and soft contributions. Fragmentation functions appear in this
context.
An integral part of this analysis is the
use of renormalization group equations
-SUSY DGLAP-
which resum the logarithmic scaling violations and evolve distributions
functions
and fragmentation functions to the appropriate factorization scale.

In our approach R-parity is conserved.
However, we neglect contribution from fragmentation into stable
supersymmetric states which are subleading relative to fragmentation into
strongly interacting states. Similarly, we do not consider the possibility
of R-parity violating operators. Limits on such operators \cite{driener}
would suggest that they cannot produce sizable effect on the fragmentation
function. However, more detailed studies may be required to ascertain
this general expectation.


 These studies require the matrix of the anomalous dimensions (for all the
moments), or of the corresponding Altarelli-Parisi (DGLAP) kernels.
We recall that the leading order anomalous dimensions are known
\cite{Antoniadis,KR} both
for a partial SUSY evolution and for the exact one.

\subsection{The Timelike Evolution}
In this work we focus our attention on the evolution of
fragmentation functions in the context of exact supersymmetry,
with coupled gluons and squarks.

Fig.~1 summarizes the basic strategy of our work. The figure at the top
depicts a regular evolution (a one-phase evolution), while the one at the
bottom
illustrates a mixed QCD/SQCD evolution (a two-phase evolution).
The first stage
is supersymmetric, the second one is regular. In the first stage all the
partons
(SUSY and not SUSY) are massless, in the second one only quark and gluons
survive.
At the end of the first stage the supersymmetric partners are extinct,
but their presence at the higher scales is saved
in the boundary conditions for the
quarks and gluons fragmentation functions, when the second stage of the
evolution starts.

Within the approximation implied by the use of a pure SQCD evolution, no
missing
$E_t$ events are allowed along the development of the supersymmetric cascade,
since electroweak and flavour mixing effects are not included. Their inclusion
is still
an open chapter even in (QCD) Monte Carlo event
generators for the final state.

The ladder structure of the supersymmetric showers in the leading logarithmic
approximation can be easily pictured in terms of symmetric doubling of lines
of regular QCD ladders in all the possible allowed cases.
Showers can be equally initiated by $q\,\, \bar{q}$ pairs, gluon pairs, gluino
pairs or
squark pairs. The supersymmetric transition (i.e. those involving
supersymmetric
partners) stop (see Fig.~1 (bottom)) once the supersymmetry breaking scale
$M_{\lambda}=M_{\tilde{q}}$ is reached.

Similarly to the QCD case, in the case of exact
$N=1$ supersymmetry we define singlet and non-singlet fragmentation functions
$D^h_f(x,Q^2)$.
They describe the amplitude for a parton of type $f$
to fragment into a hadron $h$ as a function
of the Bjorken variable $x$ (fractional energy of the fragment)
and initial energy $Q$. Their operatorial definition is similar
to that of ordinary parton distributions. The evolution equations for these
functions
are similar to the standard DGLAP equations, but with transposed kernel matrix
$(P\to P^T)$. In leading order the analytic continuation (from the spacelike
or DIS evolution to the timelike evolution) is straightforward, while
some complications appear to next-to-leading order. They involve a breaking of
the Drell-Yan-Levy relation, which appears to be
violated at parton level \cite{blumlein}. Since our analysis is in the leading
logarithmic approximation,
we will not consider these aspects in our work any further.

The equations for the timelike evolution are given by

\beqa
\frac{d}{d\log(Q^2)}D^h_g(x,Q^2) &=&
\frac{\alpha_s}{2\pi}\left(P_{gg}\otimes D^h_g + P_{\lambda g}\otimes
D^h_\lambda +
P_{qg}\otimes \sum_i\left(D^h_{q_i} + D^h_{\bar{q}_i}\right)\right. \nonumber
\\
&&\left.\,\,\,\,\,\,\,\,\,\,\,\, + P_{\tilde{q}
g}\otimes\sum_{i=1}^{n_f}\left(\tilde{q}_{iL} + \tilde{q}_{iR}
+ \bar{\tilde{q}}_{iL} + \bar{\tilde{q}}_{iR}\right)\right) \\
\frac{d}{d\log(Q^2)}D^h_\lambda(x,Q^2) &=&
\frac{\alpha_s}{2\pi}\left(P_{g\lambda}\otimes D^h_g + P_{\lambda
\lambda}\otimes D^h_\lambda +
P_{q \lambda}\otimes \sum_i\left({D^h_{q_i} +
D^h_{\bar{q}_i}}\right)\right.\nonumber \\
 &&\left.\,\,\,\,\,\,\,\,\,\,\,\, + P_{\tilde{q}\lambda}\otimes
\sum_{i=1}^{n_f}\left( \tilde{q}_{iL} + \tilde{q}_{iR} + \bar{\tilde{q}}_{iL}
+ \bar{\tilde{q}}_{iR}\right)\right)\\
\frac{d}{d\log(Q^2)}D^h_{q_i}(x,Q^2) &=&
\frac{\alpha_s}{2\pi}\left(\frac{1}{2 n_f}P_{g q}\otimes D^h_g + \frac{1}{2
n_f}P_{\lambda q}\otimes D^h_\lambda +
P_{\tilde{q} q }\otimes \left( D^h_{{\tilde{q}}_{iL}} +
D^h_{{\tilde{q}}_{iR}}  \right)\right.\nonumber \\
 &&\left. \,\,\,\,\,\,\,\,\,\,\,\, + P_{qq}\otimes D^h_{{q_i}}\right)  \\
\frac{d}{d\log(Q^2)}D^h_{{\tilde{q}}_{iL}} (x,Q^2) &=&
\frac{\alpha_s}{2\pi}\left(\frac{1}{4 n_f}P_{g\tilde{q} }\otimes D^h_g +
\frac{1}{4 n_f}P_{\lambda \tilde{q}}\otimes
D^h_\lambda + \frac{1}{2}P_{q\tilde{q} }\otimes D^h_{{q}_{i}}\right.\nonumber
\\
 &&\left.\,\,\,\,\,\,\,\,\,\,\,\, + P_{\tilde{q}\tilde{q}}\otimes
D^h_{\tilde{q}_{iL}}\right) \\
\frac{d}{d\log(Q^2)}D^h_{\tilde{q}_{iR}}(x,Q^2) &=&
\frac{\alpha_s}{2\pi}\left(\frac{1}{4 n_f}P_{g\tilde{q} }\otimes D^h_g +
\frac{1}{4 n_f}
P_{\lambda \tilde{q}}\otimes
D^h_\lambda +\frac{1}{2}P_{q\tilde{q} }\otimes D^h_{q_i}\right.\nonumber \\
 &&\left.\,\,\,\,\,\,\,\,\,\,\,\, + P_{\tilde{q}\tilde{q}}\otimes
D^h_{\tilde{q}_{iR}}\right)\nonumber \\
\eeqa
In the following we will use the short-hand notation
\beq
D^h_{\tilde{q}_i}(x,Q^2)\equiv D^h_{\tilde{q}_{iL}}(x,Q^2) +
D^h_{\tilde{q}_{iR}}(x,Q^2)
\eeq
to denote the fragmentation functions of squarks of flavour $i$
at a fractional energy $x$ and momentum $Q$. It is also convenient to separate
the equations, as usual, into singlet and non singlet sectors using the
definitions

\beqa
 D^h_{q_V}(x, Q^2) &\equiv& \sum_{i=1}^{n_f}\left( D^h_{{q}_i}(x,Q^2) -
D^h_{{\bar{q}}_i}(x, Q^2)\right) , \nonumber \\
&\equiv & D^h_{q_{NS}}\nonumber \\
 D^h_{\tilde{q}_V}(x, Q^2) &=& \sum_{i=1}^{n_f}
\left( D^h_{\tilde{q}_i}(x, Q^2) - D^h_{\tilde{\bar{q}}_i}(x, Q^2)\right)
\nonumber \\
& \equiv &D^h_{\tilde{q}_{NS}}(x,Q^2)\nonumber \\
 D_{q^{+}}(x, Q^2) &\equiv& \sum_{i=1}^{n_f} \left(D^h_{{q}_i}(x, Q^2) +
D^h_{\bar{q}_i}(x, Q^2)\right) \nonumber \\
D^h_{\tilde{q}^{+}}(x, Q^2)&\equiv& \sum_{i=1}^{n_f}\left(
D^h_{\tilde{q}_i}(x,Q^2) + D^h_{\tilde{\bar{q}}_i}(x,Q^2)\right).
\eeqa

The non singlet equations are
\beqa
Q^2 {d\over d Q^2} D{q_V}(x,Q^2)&=&{\alpha(Q^2)\over 2 \pi} \left(
P_{qq}\otimes D^h_{q_V} + P_{\tilde{q}q}\otimes
D_{\tilde{q_V}}\right)  \nonumber \\
Q^2 {d\over d Q^2} D_{\tilde{q}_V}(x,Q^2)&=&{\alpha(Q^2)\over 2 \pi} \left(
P_{q\tilde{q}}\otimes D^h_{q_V} +P_{\tilde{q}\tilde{q}}\otimes
{D^h_{\tilde{q_V}}}\right),  \nonumber \\
\label{susyns}
\eeqa

and the singlet matrix equations, which mix $q_V$ and $\tilde{q}_V$ with the
gluons and the gluinos

\beq
 Q^2 {d\over d Q^2}  \left[\begin{array}{c}   D^h_g(x,Q^2)\\
D^h_{\lambda}(x,Q^2) \\ D^h_{q^+}(x,Q^2) \\D^h_{\tilde{q}^+}(x,Q^2)
\end{array} \right]
=\left[ \begin{array}{llll}
         P_{gg} & P_{g \lambda} & P_{g q} &  P_{g \tilde{q}} \\
         P_{\lambda g} & P_{\lambda \lambda} & P_{\lambda q} & P_{\lambda
\tilde{q}} \\
         P_{q g} & P_{q \lambda} & P_{q q} & P_{q \tilde{q}} \\
         P_{\tilde{q} g} & P_{\tilde{q} \lambda} & P_{\tilde{q} q} &
P_{\tilde{q} \tilde{q}}
         \end{array} \right]^T \otimes
  \left[ \begin{array}{c}
         D^h_g(x,Q^2)\\ D^h_\lambda(x,Q^2) \\ D^h_{q^{(+)}}(x,Q^2) \\
D^h_{\tilde{q}^{(+)}}(x,Q^2) \end{array} \right].
\eeq
where $``T''$ indicates the matrix transposed.
To solve for all the flavours, it is convenient to introduce the linear
combinations
\beqa
\chi_i(x,Q^2) &=& D^h_{q_i^{(+)}} -\frac{1}{n_f}D^h_{q^{(+)}} \nonumber \\
\tilde{\chi}_i(x,Q^2) &=& D^h_{\tilde{q}_i^{(+)}} -\frac{1}{n_f}
D^h_{\tilde{q}^{(+)}} \nonumber \\
\eeqa
and the additional singlet equations

\beqa
Q^2 {d\over d Q^2} D_{q_i^{(-)}}(x,Q^2)&=&
{\alpha(Q^2)\over 2 \pi} \left( P_{qq}\otimes D^h_{q_i^{(-)}} +
P_{\tilde{q}q}\otimes
D_{\tilde{q_i}^{(-)}}\right)  \nonumber \\
Q^2 {d\over d Q^2} D_{\tilde{q}_i^{(-)}}(x,Q^2)&=&
{\alpha(Q^2)\over 2 \pi} \left( P_{q\tilde{q}}\otimes D^h_{q_i^{(-)}}
+P_{\tilde{q}\tilde{q}}\otimes {D^h_{\tilde{q}_i^{(-)}}}\right)  \nonumber \\
\label{susyns1}
\eeqa
and
\beqa
Q^2 {d\over d Q^2} D^h_{\chi_i}(x,Q^2)&=&
{\alpha(Q^2)\over 2 \pi} \left( P_{qq}\otimes D^h_{\chi_i} +
P_{\tilde{q}q}\otimes
D^h_{\tilde{\chi_i}}\right)  \nonumber \\
Q^2 {d\over d Q^2} D^h_{\tilde{\chi}_i}(x,Q^2)&=&
{\alpha(Q^2)\over 2 \pi} \left( P_{q\tilde{q}}\otimes D^h_{\chi_i}
+P_{\tilde{q}\tilde{q}}\otimes D^h_{\tilde{\chi_i}}\right).  \nonumber \\
\label{susyns2}
\eeqa
The general flavour decomposition is obtained by solving the singlet equations
for
$D^h_{q^{(+)}}$ and $D^h_{\tilde{q}^{(+)}}$, then solving the non-singlet
equations
for $D^h_{q_i^{(-)}}$ and $D^h_{\tilde{q}_i^{(+)}}$ and for $D^h_{\chi_i}$
and $D^h_{\tilde{\chi}_i}$. The fragmentation
functions of the various flavours
are extracted using the linear combinations

\beqa
D^h_{q_i} &=& \frac{1}{2}\left(D^h_{q_i^{(-)}} + D^h_{\chi_i}
+\frac{1}{n_f}D^h_{q^{(+)}}\right)
\nonumber \\
D^h_{\bar{q}_i} &=& -\frac{1}{2}\left(D^h_{q_i^{(-)}} - D^h_{\chi_i}
-\frac{1}{n_f}D^h_{q^{(+)}}\right) \nonumber \\
\eeqa

and
\beqa
D^h_{\tilde{q}_i} &=& \frac{1}{2}\left(D^h_{\tilde{q}_i^{(-)}} +
D^h_{\tilde{\chi}_i}
+\frac{1}{n_f}D^h_{\tilde{q}^{(+)}}\right)
\nonumber \\
D^h_{\bar{{\tilde{q}}}_i} &=& -\frac{1}{2}\left(D^h_{\tilde{q}_i^{(-)}}
- D^h_{\tilde{\chi}_i} -\frac{1}{n_f}D^h_{\tilde{q}^{(+)}}\right) \nonumber \\
\eeqa
to identify the various flavour components.

There are simple ways to calculate the kernel of the SUSY DGLAP evolution by a
simple extension
of the usual methods.
The changes are primarily due to color factors. There are also some
basic supersymmetric relations which have to be satisfied \cite{CC1}.
They are generally broken in the case of decoupling.
We recall that the supersymmetric version of the $\beta$ function is given at
two-loop level by

\beqa
\beta_0^S &=& \frac{1}{3}\left(11\, C_A - 2\, n_f - 2 \,n_\lambda \right)
\nonumber \\
\beta_1^S &=&\frac{1}{3}\left( 34\, C_A^2 - 10\, C_A\, n_f - 10\, C_A\,
n_\lambda
- 6\, C_F\, n_f - 6 \,C_\lambda\, n_\lambda \right)
\eeqa
where $n_f$ is the number of flavours and
for Majorana gluinos $n_\lambda=C_A$.

The ordinary running of the coupling is replaced by its supersymmetric running

\beq
{\alpha^S(Q_0^2)\over 2 \pi}={2\over \beta_0^S}
{ 1\over \ln (Q^2/\Lambda^2)}\left( 1 - {\beta_1^S\over \beta_0^S}
{\ln\ln(Q^2/\Lambda^2)\over \ln(Q^2/ \Lambda^2)} +
O({1\over\ln^2(Q^2/\Lambda^2) })\right).
\eeq

The kernels are modified both in their coupling $(\alpha \to \alpha^S)$
and in their internal structure (Casimirs, color factors, etc.) when moving
from the QCD case to the SQCD case. In order
to illustrate the approach that we are going to follow in our analysis of the
fragmentation functions within a mixed SQCD/QCD evolution,
we recall the strategy employed in Refs.~\cite{CC1,CC2} to generate the
ordinary
supersymmetric distribution functions.

In Ref.~\cite{CC2} for scaling violations affecting the initial state were
introduced 3
regions:
\begin{itemize}
\item 1) the QCD region, described by ordinary QCD (Altarelli Parisi or AP);

\item 2) an intermediate supersymmetric region with coupled gluinos
and decoupled squarks (partially supersymmetric AP or SAP);

\item
3) the N=1 region (exact supersymmetric AP or ESAP).
\end{itemize}
The basic idea was to generate these distributions {\em radiatively}
as is usually
done in QCD for the gluons, for instance, using the fact that the
matrix of the anomalous
dimensions is not-diagonal.
The most general sequence of evolutions
(denoted AP-SAP-ESAP in \cite{CC2}) is described by the arrays
$(Q_i,Q_f)_{AP}-(Q_i,Q_f)_{SAP}-(Q_i,Q_f)_{ESAP}$, with
$Q_{f,AP}=Q_{i,SAP}=m_{2\lambda}$
and $Q_{f,SAP}=Q_{i,ESAP}=m_{\tilde{q}}$. In this work we limit our
analysis to a
simpler AP-ESAP evolution and we run the DGLAP equations {\em back}
starting
from known QCD fragmentation functions - for which
various sets are available in the literature \cite{kkp} -
up to an intermediate supersymmetric scale. From this point on (up in energy)
we switch
on the ESAP evolution of fragmentation functions. As we run the equations
upward,
supersymmetric fragmentation functions are generated. The boundary values of
the
low energy (QCD) functions at the supersymmetry scale $m_{2\lambda}$ set
the initial condition  for the (backward) supersymmetric running
up to the final scale. In simple terms: we reach the mountain
from the valley, and cross a fence along the way.

In general, if we split the intermediate SUSY scale into 2 sectors
$m_{\lambda} << m_{\tilde{q}}$ (ESAP-SAP-AP evolution)
in this (general) case the solution is built by sewing the three regions as

\beqa
\left[ \begin{array}{c}
         D^h_{q_V}(x,Q^2)\\
         D^h_{\tilde{q}_V}(x,Q^2)        \\
         \end{array} \right] &=& \left[ \begin{array}{c}
         D^h_{q_V}(x,Q_0^2)\\
         0        \\
         \end{array} \right] + \int_{Q_0^2}^{m_{2\lambda}^2} d\, \log\, Q^2\,
P_{AP}^{T,NS}(x,\alpha(Q^2))\otimes \left[ \begin{array}{c}
         D^h_{q_V}(x,Q^2)\\
         0        \\
         \end{array} \right] \nonumber \\
&+& \int_{m_{2\lambda}^2}^{m_{ 2\tilde{q}}^2} d\, \log Q^2
P_{SAP}^{T,NS}(x,\alpha^S(Q^2))\otimes \left[ \begin{array}{c}
         D^h_{q_V}(x,Q^2)\\
         0        \\
         \end{array} \right] \nonumber \\
&+& \int_{m_{2\lambda}^2}^{Q_f^2} d\, \log Q^2\,
P_{ESAP}^{T,NS}(x,\alpha^{ES}(Q^2))\otimes \left[ \begin{array}{c}
         D^h_{q_V}(x,Q^2)\\
         D^h_{2\tilde{q}^V}(x,Q^2)        \\
         \end{array} \right] \nonumber \\
\eeqa
in the non singlet and

\beqa
\left[ \begin{array}{c}
         D^h_{G}(x,Q_f^2)\\
	 D^h_{\lambda}(x,Q^2)\\
	 D^h_{q^+}(x,Q^2)\\
         D^h_{\tilde{q}^+}(x,Q^2)\\
         \end{array} \right] &=& \left[ \begin{array}{c}
         D^h_G(x,Q_f^2)\\
	 0\\
	  D^h_{q^+}(x,Q_f^2) \\
         0\\
         \end{array} \right]  + \int_{Q_0^2}^{m_{2\lambda}^2} d\, \log\, Q^2\,
P_{T,AP}^{NS}(x,\alpha(Q^2))\otimes
\left[ \begin{array}{c}
         D^h_G(x,Q_f^2)\\
	  0\\
	 D^h_{q^+}(x,Q^2)\\
         0\\
         \end{array} \right] \nonumber \\
&+& \int_{m_{2\lambda}^2}^{m_{2 \tilde{q}}^2} d\, \log Q^2
P_{T,SAP}^{NS}(x,\alpha^S(Q^2))\otimes
\left[ \begin{array}{c}
         D^h_G(x,Q_f^2)\\
	 D^h_{\lambda}(x,Q^2)\\
	 D^h_{q^+}(x,Q^2)\\
           0\\
         \end{array} \right] \nonumber \\
&+& \int_{m_{2\lambda}^2}^{Q_f^2} d\, \log Q^2\,
P_{T,ESAP}^{NS}(x,\alpha^{ES}(Q^2))\otimes
\left[ \begin{array}{c}
         D^h_G(x,Q_f^2)\\
	 D^h_{\lambda}(x,Q^2)\\
	 D^h_{q^+}(x,Q^2)\\
           D^h_{2\tilde{q}^+}(x,Q^2)\\
         \end{array} \right] \nonumber \\
\eeqa
for the singlet solution, where $m_{2\tilde{q}}=2\, m_{\tilde{q}}$.
The zero entries in the arrays for some of the distributions are due
to the boundary conditions, since all the supersymmetric partners are
generated, in this model,
by the evolution.

The general structure of the algorithms that solves these equations
is summarized below. We start from the non singlet sector
and then proceed to the singlet.
{}From now on, to simplify our notation, we will omit at times the
index ``$h$'' from the fragmentation functions when obvious.
We define
$D_{q^{NS}}(x,Q^2)=\left(D_{q_V}(x,Q^2),D_{\tilde{q}_V}(x,Q^2)\right)^T$  and
$DA_n(x)=\left(DA^{q_V}_n,DA^{\tilde{q}_V}_n\right)^T$
and introduce the ansatz

\beq
D_{q^{NS}}(x,Q^2)=\sum_{n=0}^{n_0}\,{A_n(x)\over n!} \log^n
\left(\frac{\alpha(Q^2)}{\alpha{(Q_0^2)}}\right),
\eeq
where $n_0$ is an integer at which we stop the iteration, which
usually ranges up to 20.
The first coefficient of the recursion is determined by the initial condition
\beq
DA_0(x) = U(x)\otimes D_q^{NS}(x,Q_0^2),
\eeq
where
\beqa
U_1(x) &\equiv &
\left[ \begin{array}{ll}
         \delta(1-x) & 0\\
	  0 & 0\\
         \end{array} \right] \nonumber \\
U_{1\,2}(x) &\equiv &
\left[ \begin{array}{ll}
         \delta(1-x) & 0\\
	  0 & \delta(1-x)\\
         \end{array} \right].
\eeqa

The recursion relations are given by

\beq
DA_{n+1}(x)= -\frac{2}{\beta_0}P^{T,NS}_{AP}\otimes DA_n(x)
\eeq

The solution in the first (DGLAP) region at the first matching scale
$m_{2\lambda}$ is given by

\beq
D_{q^{NS}}(x,m_{2\lambda})=\sum_{n=0}^{n_0}\frac{DA^{AP}_n(x)}{n!}\log^n
\left(\frac{\alpha({m_{2\lambda}}^2)}{\alpha(Q_0^2)}\right).
\eeq
valid for $m_{\lambda} < m_{\tilde q}$. As we will discuss
in the next sections,
these two scales will be taken to be degenerate and the numerical
treatment simplifies considerably. We also mention that we
don't find any significant difference,
from our analysis, if this degeneracy between the gluino and squark masses
is lifted. However, just for illustrative purposes, we keep the two scales
separated in the formal analysis presented below in order to show
how the matching is performed in a more general case. Mass differences
within a range of 100-200 GeV play a rather modest role in the analysis
and are hardly distinguishable.

At the second stage the (partial) supersymmetric coefficients are given by
(S is a short form of $SAP$)

\beqa
DA^{S,NS}_0(x) &=& U(x)\otimes D_q(x,m_{2\lambda^2}) \nonumber \\
DA^{S,NS}_{n+1}(x) &=& -\frac{2}{\beta^S_0} P^{T,NS}_{S}(x)\otimes
DA^{S,NS}_n(x).\nonumber \\
\eeqa

We construct  the boundary condition for the next stage of the evolution using
the intermediate solution
\beq
D_q(x,Q^2)=\sum_{n=0}^{n_0}\frac{DA^S_n(x)}{n!}
\left(\frac{\alpha({Q^2})}{\alpha({m_{2\lambda}}^2)}\right).
\eeq

evaluated at the next threshold $m_{2 \lambda}$.

\beq
D_q(x,Q^2)=\sum_{n=0}^{n_0}\frac{DA^S_n(x)}{n!}
\log\left(\frac{\alpha({Q^2})}{\alpha({m_{2\lambda}}^2)}\right).
\eeq

The final solution is constructed using the recursion relations

\beqa
DA^{ES}_0(x) &=& U(x)\otimes q(x,m_{ \tilde{q}}) \nonumber \\
DA^{ES}_{n+1}(x) &=& -\frac{2}{\beta^{ES}}P^{NS}_{ES}(x)\otimes DA^{ES}_n(x)
\nonumber \\
\eeqa

The final solution is written as
\beqa
D_{q^{NS}}(x,Q^2) &=& D_{q^{NS}}(x,Q_0^2) + \sum_{n=1}^{n_0}
\frac{DA_n(x)}{n!}\log\left(\frac{\alpha({m_{2\lambda}}^2)}
{\alpha(Q_0^2)}\right)+
\sum_{n=1}^{n_0}\frac{DA^{S}_n(x)}{n!}\log
\left(\frac{\alpha^{S}(m_{\tilde{q}}^2)}
{\alpha^{S}({m_{2\lambda}}^2)}\right) \nonumber \\
& + &
\sum_{n=1}^{n_0}\frac{DA^{ES}_n(x)}{n!}\log\left(\frac{\alpha^{ES}(Q_f^2)}
{\alpha^{ES}({m_{\tilde{q}}}^2)}\right) \nonumber \\
\eeqa

As we have already mentioned above, in the analysis presented below the two
SUSY scales
will be collapsed into one ($m_{2\lambda}$).

\section{QCD Evolution of the Fragmentation Functions}
As we have already mentioned, the timelike and the spacelike evolution, in
leading order, are essentially the same. The singlet kernels
are just the transposed of the kernels describing the evolution of the
parton distributions. The ordinary QCD kernels, in leading order, are given by

\beqa
P^{(0)}_{qq, NS} &=& C_F\left(\frac{1 + x^2}{1-x}\right)_+\nonumber \\
&=& C_F\left( \frac{2}{(1-x)_+}- 1 - x + \frac{3}{2}\,\delta(1-x)\right)
\nonumber \\
P^{(0)}_{qq}(x)&=& P^{(0)}_{qq,NS}\nonumber \\
P^{(0)}_{qg}(x)&=&  \,\,n_f \left( x^2 + (1-x)^2\right)\nonumber \\
P^{(0)}_{gq}(x)&=& C_F {1 + (1-x)^2\over x}\nonumber \\
P^{(0)}_{gg}(x)&=& 2 N_c\left( {1\over (1-x)_+} + {1\over x}
-2 + x (1-x)\right) + {\beta_{o}\over 2}\delta(1-x)
\eeqa

with $\beta_0=11/3 \,C_A -4/3\, T_R\, n_f$ being
the first coefficient of the QCD $\beta$-function and $T_R=1/2$. $n_f$ is the
numbers of
flavours.
The equations for the fragmentation functions in QCD are given by
\beqa
\frac{d}{d \log(Q^2)}D^h_{q_i}(x,Q^2)
&=&\frac{\alpha(Q^2)}{2 \pi}
\left( P_{qq}\otimes D^h_{q_i}+ \frac{1}{2 n_f}P_{g q}\otimes D^h_g\right)
\nonumber \\
\frac{d}{d \log(Q^2)}D^h_{g}(x,Q^2)
&=&\frac{\alpha(Q^2)}{2\pi}
\left( P_{qg}\otimes \sum_i\left(D^h_{q_i}+D^h_{\bar{q}_i}\right)
+ P_{g g}\otimes D^h_g \right)\nonumber \\
\eeqa
and, as usual, can be decomposed into a non-singlet and a singlet sector

\beqa
Q^2 {d\over d Q^2} D^h_{q_i^{(-)}}(x, Q^2) &=&
{\alpha(Q^2)\over 2 \pi} P_{qq,\,NS}(x, \alpha(Q^2))\otimes
D^h_{q_{i}^{(-)}}(x, Q^2)
\eeqa
for the non-singlet distributions and
\beqa
&&  { d\over d \log(Q^2)} \left( \begin{array}{c}
D^h_{q^{(+)}}(x, Q^2) \\
D^h_g(x, Q^2) \end{array}\right)=
\left(\begin{array}{cc}
P_{qq} & P_{gq} \\
P_{qg} & P_{gg}
\end{array} \right)\otimes
\left( \begin{array}{c}
D^h_{q^{(+)}}(x,Q^2) \\
D^h_g(x,Q^2) \end{array}\right) \nonumber \\
\eeqa
for the singlet sector, where

\beq
D^h_{q^{(-)}_i}(x,Q^2)=D^h_{q_i}- D^h_{\bar{q}_i}.
\eeq
A fast strategy to solve these equations, as discussed in \cite{CC1},
where a complete leading order evolution has been implemented,
is to solve the recursion relations
numerically. In the timelike case -that we are considering-
these relations are obtained
as linear combinations of the spacelike ones. We get

\beqa
DA_{n+1}^{q_i^{(-)}}&=&-4 \frac{C_F}{\beta_0}\int_x^1 \frac{dy}{y}
\frac{y DA_n^{q_i^{(-)}}(y) - x DA_n^{q_i^{(-)}}(x) }{y-x} -
4 \frac{C_F}{\beta_0} \log(1-x)DA_n^{q_i^{(-)}}(x)\nonumber \\
&& + 2\frac{C_F}{\beta_0}\int_x^1\frac{dy}{y}\left(1 +
z\right)DA_n^{q_i^{(-)}}(y) -
3\frac{C_F}{\beta_0}DA_n^{q_i^{(-)}}(x)
\nonumber \\
DA_{n+1}^{q^{(-)}}&=&-4 \frac{C_F}{\beta_0}\int_x^1 \frac{dy}{y}
\frac{y DA_n^{q^{(-)}}(y) - x DA_n^{q^{(-)}}(x) }{y-x} -
4 \frac{C_F}{\beta_0} \log(1-x)DA_n^{q^{(-)}}(x)\nonumber \\
&& + 2\frac{C_F}{\beta_0}\int_x^1\frac{dy}{y}\left(1 +
z\right)DA_n^{q^{(-)}}(y) -
3\frac{C_F}{\beta_0}DA_n^{q^{(-)}}(x). \nonumber \\
\eeqa
A similar expansion is set up for the non-singlet variable
$\chi_i=q_i^{(+)} - 1/n_f\,\, q^{(+)}$
\beqa
DA_{n+1}^{\chi_i}&=&-4 \frac{C_F}{\beta_0}\int_x^1 \frac{dy}{y}
\frac{y DA_n^{\chi_i}(y) - x DA_n^{\chi_i}(x) }{y-x} -
4 \frac{C_F}{\beta_0} \log(1-x)DA_n^{\chi_i}(x)\nonumber \\
&& + 2\frac{C_F}{\beta_0}\int_x^1\frac{dy}{y}\left(1 + z\right)
DA_n^{\chi_i}(y) -
3\frac{C_F}{\beta_0}DA_n^{\chi_i}(x). \nonumber \\
\eeqa
Similarly, the singlet equations generate recursion relations of the form

\beqa
DA_{n+1}^{q^{(+)}}(x) &=& -4\frac{C_F}{\beta_0}\int_x^1\frac{dy}{y}
\frac{y DA_n^{q^{(+)}}(y) - x DA_n^{q^{(+)}}(x)}{y-x} - 4
\frac{C_F}{\beta_0}\log(1-x)
DA_n^{q^{(+)}}(x)\nonumber \\
&& + 2\frac{C_F}{\beta_0}\int_x^1\frac{dy}{y}\left( 1+
z\right)DA_n^{q^{(+)}}(y)
-3\frac{C_F}{\beta_0}DA_n^{q^{(+)}}(x)
\nonumber \\
&& -2\frac{C_F}{\beta_0 }\int_x^1\frac{dy}{y}
\frac{1 +(1-z)^2}{z}DA_n^g(y)
\nonumber \\
DA_{n+1}^g(x) &=&-4 \frac{C_A}{\beta_0}\int_x^1\frac{dy}{y}
\frac{y DA_n^{g}(y) - x DA_n^{g}(x)}{y-x} - 4\frac{C_A}{\beta_0}
\log(1-x)DA_n^g(x)\nonumber \\
&& - 2\frac{n_f}{\beta_0}\int_x^1\frac{dy}{y}\left( z^2
+(1-z)^2\right)DA_n^{q^{(+)}}(y)
- D A_n^g(x)\nonumber \\
&& -4 \frac{C_A}{\beta_0}\int_x^1\frac{dy}{y}
\left( \frac{1}{z} -2 +z(1-z)\right)DA_n^g(y).
\eeqa

The coefficients of the various flavours both for $D^h_{q_i}$ and
$D^h_{\bar{q}_i}$
are then obtained from the relations

\beqa
DA_n^{q_i}&=&\frac{1}{2}\left( DA_n^{\chi_i} +
\frac{1}{n_f}DA_n^{q_i^{(-)}}\right)
\nonumber \\
DA_n^{\bar{q}_i}&=&\frac{1}{2}\left( DA_n^{\chi_i} -
\frac{1}{n_f}DA_n^{q_i^{(-)}}\right)
\eeqa

\section{ The ESAP ($N=1$ QCD) Fragmentation}

Moving to $N=1$ QCD, we introduce recursion relations for appropriate
linear combinations of non-singlet fragmentation functions

\beqa
DA_{n+1}^{\chi_i} &=& -4\frac{C_F}{\beta_0}\int_x^1\frac{dy}{y}
\frac{y DA_n^{\chi_i}(y) - x DA_n^{\chi_i}(x)}{y-x}
- 4 \frac{C_F}{\beta_0}\log(1-x) DA_n^{\chi_i}(x)
\nonumber \\
&& - 2\frac{C_F}{\beta_0}DA_n^{\chi_i}(x)
- 2\frac{C_F}{\beta_0}\int_x^1\frac{dy}{y}z DA_n^{\tilde{\chi}_i}(y)
+2\frac{C_F}{\beta_o}
\int_x^1\frac{dy}{y}\left(1 +z\right)D A_n^{\chi_i}(y) \nonumber \\
\eeqa

\beqa
DA_{n+1}^{\tilde{\chi}_i} &=& -4\frac{C_F}{\beta_0}\int_x^1\frac{dy}{y}
\frac{y DA_n^{\tilde{\chi}_i}(y) - x DA_n^{\tilde{\chi}_i}(x)}{y-x}
- 4 \frac{C_F}{\beta_0}\log(1-x) DA_n^{\tilde{\chi}_i}(x) \nonumber \\
&& - 2\frac{C_F}{\beta_0}DA_n^{\tilde{\chi}_i}(x)
- 2\frac{C_F}{\beta_0}\int_x^1\frac{dy}{y}z DA_n^{\chi_i}(y)
+2\frac{C_F}{\beta_o}
\int_x^1\frac{dy}{y}\left(1 +z\right)D A_n^{\tilde{\chi_i}}(y)\nonumber \\
\eeqa

and two similar non-singlet equations for $D^h_{q_i^{(-)}}$ and
$D^h_{\tilde{q}_i^{(-)}}$

\beqa
DA_{n+1}^{q_i^{(-)}} &=& -4\frac{C_F}{\beta_0}\int_x^1\frac{dy}{y}
\frac{y DA_n^{q^{(-)}_i}(y) - x DA_n^{q^{(-)}_i}(x)}{y-x}
- 4 \frac{C_F}{\beta_0}\log(1-x) DA_n^{q^{(-)}_i}(x) \nonumber \\
&& - 2\frac{C_F}{\beta_0}DA_n^{q^{(-)}_i}(x)
- 2\frac{C_F}{\beta_0}\int_x^1\frac{dy}{y}z DA_n^{\tilde{q}^{(-)}_i}(y)
+2\frac{C_F}{\beta_o}
\int_x^1\frac{dy}{y}\left(1 +z\right)D A_n^{q_i^{(-)}}(y)
 \nonumber \\
\label{one}
\eeqa

\beqa
DA_{n+1}^{\tilde{q}^{(-)}_i} &=& -4\frac{C_F}{\beta_0}\int_x^1\frac{dy}{y}
\frac{y DA_n^{\tilde{q}^{(-)}_i}(y) - x DA_n^{\tilde{q}^{(-)}_i}(x)}{y-x}
- 4 \frac{C_F}{\beta_0}\log(1-x) DA_n^{\tilde{q}^{(-)}_i}(x) \nonumber \\
&& - 2\frac{C_F}{\beta_0}DA_n^{\tilde{q}^{(-)}_i}(x)
- 2\frac{C_F}{\beta_0}\int_x^1\frac{dy}{y}z DA_n^{q^{(-)}_i}(y)
+2\frac{C_F}{\beta_o}
\int_x^1\frac{dy}{y}\left(1 +z\right)D A_n^{\tilde{q}_i^{(-)}}(y).\nonumber \\
\label{two}
\eeqa

Analogous equations are satisfied by the combinations
$D^h_{q^{(-)}}$ and $D^h_{\tilde{q}^{(-)}}$ by replacing in (\ref{one}) and
(\ref{two})
$D^h_{q_i^{(-)}}\to D^h_{q^{(-)}}$ and $D^h_{\tilde{q}_i^{(-)}} \to
D^h_{\tilde{q}^{(-)}}$.

Moving to the singlet sector we obtain
\beqa
DA_{n+1}^g(x) &=& - 4 \frac{C_A}{\beta_0}\int_x^1 \frac{dy}{y}
\frac{y DA_n^g(y) - x DA_n^g (x)}{ y -x}
- 4 \frac{C_A}{\beta_0}\log(1-x)DA_n^g(x) -D A_n^g(x)
\nonumber \\
&& + 2 \frac{C_A}{\beta_0}\int_x^1\frac{dy}{y}\left( 1 + z\right)DA_n^g(y)
 -2 \frac{C_A}{\beta_0}\int_x^1\frac{dy}{y}\left(
\frac{2}{z} + z - 2\right)DA_n^g(y)\nonumber \\
&& +  2 \frac{C_A}{\beta_0}\int_x^1\frac{dy}{y}\left(z^2 + (1-z)^2\right)
DA_n^g(y)
 -2 \frac{n_f}{\beta_0}\int_x^1\frac{dy}{y}
\left( z^2 +(1-z)^2\right) DA_n^q(y)\nonumber \\
&& -2 \frac{C_A}{\beta_0}\int_x^1\frac{dy}{y}\left(z^2 +
(1-z)^2\right)DA_n^\lambda(y)
-4\frac{n_f}{\beta_0}\int_x^1\frac{dy}{y}\, z\left( 1-z\right)
DA_n^{\tilde{q}}(y) \nonumber \\
\eeqa

\beqa
DA_n^{\lambda}(x) &=& -2\frac{C_A}{\beta_0}\int_x^1
\frac{dy}{y}\left(\frac{2}{z} + z - 2\right)DA_n^g(y)
-4 \frac{C_A}{\beta_0}\int_x^1\frac{dy}{y}
\frac{DA_n^\lambda(y)- x DA_n^\lambda(x)}{y-x}\nonumber \\
&& -4 \frac{C_A}{\beta_0}\log(1-x)DA_n^\lambda(x)
+2 \frac{C_A}{\beta_0}\int_x^1 \frac{dy}{y}
\left( 1 +z\right)DA_n^\lambda(y) - DA_n^\lambda(x)
\nonumber \\
&& -2\frac{n_f}{\beta_0}\int_x^1\frac{dy}{y}\left( 1-z \right)DA_n^q(y)
-2\frac{n_f}{\beta_0}\int_x^1\frac{dy}{y} z DA_n^{\tilde{q}}(y)
\eeqa

\beqa
DA_{n+1}^{q^{(+)}}(x) &=& -2\frac{C_F}{\beta_0}\int_x^1
\frac{dy}{y}\left(\frac{2}{z} + z -2\right)DA_n^g(y)
-2\frac{C_F}{\beta_0}\int_x^1\frac{dy}{y}\left( 1-z \right)
DA_n^\lambda(y)\nonumber \\
&& -4 \frac{C_F}{\beta_0}\int_x^1\frac{dy}{y}
\frac{y DA_n^{q^{(+)}}(y) - xDA_n^{q^{(+)}}(x)}{y-x} -
4\frac{C_F}{\beta_0} \log(1-x) DA_n^{q^{(+)}}(x)
\nonumber \\
&&
+ 2 \frac{C_F}{\beta_0}\int_x^1 \frac{dy}{y}
\left( 1 + z\right)DA_n^{q^{(+)}}(y) -2\frac{C_F}{\beta_0}DA_n^{q^{(+)}}(x)
-2\frac{C_F}{\beta_0}\int_x^1\frac{dy}{y}z DA_n^{\tilde{q}^{(+)}}(y)
\nonumber \\
\eeqa

\beqa
DA_{n+1}^{\tilde{q}^{(+)}}(x)&=&-2\frac{C_F}{\beta_0}\int_x^1
\frac{dy}{y}\left(\frac{2}{z}-2\right)DA_n^g(y)
-2\frac{C_F}{\beta_0}\int_x^1\frac{dy}{y}\left(DA_n^\lambda(x)
-DA_n^{\tilde{q}^{(+)}}(y)\right)
\nonumber \\
&& -4\frac{C_F}{\beta_0}\int_x^1\frac{dy}{y}
\frac{y DA_n^{\tilde{q}}(y) - x DA_n^{\tilde{q}}(x)}{y-x}
+ 4\frac{C_F}{\beta_0}\int_x^1\frac{dy}{y}DA_n^{\tilde{q}}(y)\nonumber \\
&& -
4\frac{C_F}{\beta_0}\log(1-x)DA_n^{\tilde{q}^{(+)}}(y)
-2\frac{C_F}{\beta_0}DA_n^{\tilde{q}^{(+)}}(x)
\eeqa
where $z\equiv x/y$.

\section{Numerical Results}
As an illustration of the procedure we adopt in our studies,
let us consider the
decay
of a hypothetical massive state into supersymmetric partons.
The decay can proceed, for instance, through a regular $q\bar{q}$
channel and a shower is developed starting from the quark pair.
The $N=1$ DGLAP
equation describes in the leading logarithmic approximation the evolution
of the shower which accompanies the pair, and we are interested in studying
the
impact of the supersymmetry breaking scale ($m_\lambda$) on the fragmentation.
In our runs we have chosen the initial set of Ref.~\cite{kkp}, with
parameterizations that can be found in the appendix.

 In our analysis we focus on the proton fragmentation
functions and, following Ref.~\cite{kkp}, we introduce the scaling variable
\begin{equation}
\bar s=\ln\frac{\ln\left(\mu^2/\Lambda_{\overline{\mathrm{MS}}}^{(5)}\right)}
{\ln\left(\mu_0^2/\Lambda_{\overline{\mathrm{MS}}}^{(5)}\right)}.
\end{equation}
In LO  we have $\Lambda_{\overline{\mathrm{MS}}}^{(5)}=88$~MeV
\cite{kkp} and use three different values for $\mu_0$, namely \cite{kkp}
\begin{eqnarray}
\mu_0&=&\left\{\begin{array}{l@{\quad\mbox{if}\quad}l}
\sqrt2~\mbox{GeV}, & a=u,d,s,g\\
m(\eta_c)=2.9788~\mbox{GeV}, & a=c\\
m(\Upsilon)=9.46037~\mbox{GeV}, & a=b
\end{array}\right..
\end{eqnarray}
This leads to three different definitions of $\bar s$.
For definiteness, we use the symbol $\bar s_c$ for charm and $\bar s_b$ for
bottom along with $\bar s$ for the residual partons.
We parameterize the f.f.'s as
\begin{equation}
\label{temp}
D(x,\mu^2)=Nx^\alpha(1-x)^\beta\left(1+\frac{\gamma}{x}\right)
\end{equation}
and express the coefficients $N$, $\alpha$, $\beta$, and $\gamma$ as
polynomials in $\bar s$, $\bar s_c$, and $\bar s_b$.
For $\bar s=\bar s_c=\bar s_b=0$, the parameterizations agree with Eq.~(2) of
Ref.~\cite{kkp} in combination with the appropriate entries in Table~2 of that
paper.
The charm and bottom parameterizations must be put to zero by hand for
$\bar s_c<0$ and $\bar s_b<0$, respectively.

Typical fragmentation functions in QCD involve final states with
$p$, $\bar{p}$, $\pi^{\pm},\pi^{0}$ and kaons $k^{\pm}$
and we refer to the original literature for a list of all the fragmentation
sets.
 We have chosen
an initial evolution scale of $10$ GeV and varied both the mass of the SUSY
partners
(we assume for simplicity that these are all degenerate)
and the final evolution scale ($Q_f$) in our backward evolution.
At the end of the evolution, once the supersymmetric fragmentation functions
are
built, we interpret $Q_f$ to be the energy available for the decay of the
massive state. We perform two types of investigations: 1) we analize the
dependence of
the evolution on the gluino mass $m_{\lambda}$, assumed to be degenerate with
the quark mass; 2) we investigate the variation in size of the fragmentation
functions
in terms of the scale $Q_f$.
Since R-parity is conserved, the fragmentation scale is $m_{2\lambda}$, two
times
the gluino mass.
Since our concern is in establishing the impact of
supersymmetric evolution and compare it to standard QCD evolution across large
evolution intervals, we plot the initial fragmentation functions,
the regularly evolved QCD functions and the SQCD/QCD evolved ones.
The latter two are originated from the same low
energy form of Ref.~\cite{kkp}.

It is possible to include in the evolution of the fragmentation
functions also different thresholds associated
with more complex spectra in which the SUSY partners are not degenerate.
These effects are negligible. Also, threshold enhancements
may require a more accurate treatment and will be discussed elsewhere. On
general grounds, however, we don't expect them to play any important role,
especially since we are interested in very extended renormalization group
runnings.

Let us now come to a description of our results.
Fig.~2 shows the initial condition for the up quark
fragmentation function into
protons, and we have chosen the initial evolution scale ($Q_0)$ to be 10 GeV,
as it has been extrapolated from collider data in Ref.~\cite{kkp}.
The evolution of this function follows QCD
from this lowest scale up to a scale of $200$ GeV $(m_{2\lambda}\equiv Q_i)$,
above which we use the full $N=1$ evolution. In this
figure we have chosen a lower initial fragmentation scale ($Q_f$) of 1 TeV.
As one can observe from the two plots for the up and down quarks,
in general, the small-x (diffractive)
region gets slightly enhanced when SUSY effects are taken into account,
although the changes are quite small.
The reason for this behaviour is  to be found in the fact that the
highest scale is not large enough to allow a
discrimination of SUSY effects from the non supersymmetric ones, since not
enough
room is available for the supersymmetric evolution.

In Fig.~3 an analogous behaviour
is found for the fragmentation functions of charm and strange quarks. Again,
small-x enhancements are seen, but regular and SUSY evolution are hardly
distinguishable. The situation appears to be quite different
for the gluon fragmentation functions (f.f's) (Fig.~4). The regular and the
SQCD evolved f.f.'s differ noticeably in the diffractive region, with a gluon
fragmenting
much faster when SUSY is in place, compared to the non supersymmetric case.
This may be related to the wider phase space available for gluons to decay
in the presence of supersymmetric channels. $Q_f$ is not too large, in this
case,
and equal to 1 TeV, which renders the plot particularly
interesting from the experimental viewpoint.

However, it should be kept in mind that even if the evolution
predicts a larger probability for the fragmentation
of a given parton into a specific hadron -compared to QCD-, this change does
not automatically translates into a noticeable effect
on the final multiplicity
of that hadron in the final state. In fact, as we are going
to show in the next section, at least in the case of the multiplicities,
SUSY effects remain small up to pretty large fragmentation scales.

As we raise the final evolution scale
we start seeing some interesting features
of the supersymmetric distributions.
For instance, this is illustrated in Fig.~5 where we show
the squark f.f.'s for all the flavours and
the one of the gluino for comparison at a nominal fragmenting initial scale
of 1 TeV. The scalar charm distribution appear
to grow slightly faster than the remaining scalar ones, but the gluino f.f.
is still the fastest growing at small-x values, which indicates
a larger proability for gluino to fragment compared to scalar quarks.

Overall, these probabilities, even at small-x values, remain small by a factor
of 100 compared to the
ordinary QCD fragmentation functions. This is not a big surprise,
given the low
scale used in this specific example for $Q_f$.

 In Figs.~6 and 7 we come to the second part of our analysis,
and we start investigating the dependence of the fragmentation functions
on a varying initial scale $Q_f$. We have selected the up quark and the gluon
in these figures. The impact of a varying scale on these functions
appear to be
rather small in the quark case, while is more pronounced for the gluon case.
Supersymmetry is broken and restored, in these two examples, at a scale of 200
GeV.

An important point which deserves thorough attention is the study of the
dependence of the fragmentation functions on the SUSY scale $m_\lambda$, which
is shown in Fig.~8. We have varied this scale in a considerable range
(from 200 GeV up to 1 TeV) and kept the initial fragmentation scale
$Q_f$ fixed at 100 TeV.
The changes for the up quark fragmentation functions are all within a
5 $\%$ interval.
A similar result is presented in Fig.~9, where we illustrate in a same plot
the dependence on the SUSY scale $m_{\lambda}$ of the fragmentation
functions of both the squark up and the up quark.

The shape of the gluino density
as a function of the the initial fragmentation scale is shown in Fig.~10.
The growth appears to be rather pronounced at small-x.
Increasing the fragmenting scale $Q_f$, fragmentation occurs at larger
x-values.

Fig.~11 illustrates the behaviour of f.f's of all types at a large initial
fragmentation scale. In order to illustrate the different behaviours
that the quark/squark sector has compared to the gluon/gluino sector,
we have shown the f.f.'s for the $b$ quark and the $\tilde{b}$ squark.
Although all the functions get a small-x enhancement or are more supported in
this
x-region, the largest fragmentation appears to come from the gluon-gluino
sector.
The gluon fragmentation function
is still the largest and fastest growing at small-x, due to the small-x
behaviour
of the gluon anomalous dimensions, while the fragmentation functions for the
bottom quark and for its superpartner are rather small.

Finally Fig.~12 has been included in order to clarify some issues concerning
the possible impact of our results on the multiplicities
of the final state hadrons due to the supersymmetric evolution.
While a more detailed analysis of this and of other similar issues will be
presented
elswewhere, here we would like to make some comments in this directions.

For this purpose,
consider the decay of the metastable state into $q\bar{q}$ pairs (for QCD) and
into a $q\bar{q}$ or a $\tilde{q}\bar{\tilde{q}}$ pairs (for $SQCD$ )
mediated by a heavy photon ($\gamma^*$). In one case we follow a
standard DGLAP evolution, in the second case we analize the
probability for the supersymmetric partons to end up into protons using
the SUSY evolution. In Fig.~12 we focus on the following observables,
characterizing the final state

\beq
R_{QCD}^h(Q^2)=\sum _{i=1}^{n_f}e_i^2\int_{z_{min}^h}^1
d\,z\left(D_{q_i}^h(z,Q^2)
+ D_{\bar{q}_i}^h(z,Q^2)\right)
\eeq

and

\beq
R_{SQCD}^h(Q^2)=\sum _{i=1}^{n_f}e_i^2\int_{z_{min}^h}^1
d\,z\left(D_{q_i}^h(z,Q^2) + D_{\bar{q}_i}^h(z,Q^2)
+ D_{\tilde{q}_i}^h(z,Q^2) + D_{\bar{\tilde{q}}_i}^h(z,Q^2)\right)
\eeq
with $ z_{min}^h= m_h/(Q/2)$ being the minimum fractional energy required for
the
fragmentation to take place. $e_i$ are the charges of the $n_f$ quark and
squark flavours,

We recall that since $D_{parton}^{h}(x,Q^2)$ is the probability
for a given parton to fragment into a final state
hadron $h$ at a given fractional energy $x$, given an original fragmentation
scale $Q$, the two observables above describe the total probability for
producing a hadron $h$, through quark channels
and their corresponding supersymmetric counterparts.

We study
the numerical values of $R_{QCD}(Q)$ and $R_{SQCD}(Q)$ as a function of Q (in
GeV), the initial fragmenting scale of partons into protons.
We show in fig.~12 the variation of $R_{QCD}$ and $R_{SQCD}$
with the initial energy scale $Q$ in a given interval (1 TeV - 4 TeV). The
fragmentation probabilities, in this specific case, differ by approximately 2
$\%$.
The opening of supersymmetric channel, in our results, tends to slightly lower
the probabilities for fragmentation
-compared to standard QCD-. Here we have chosen a nominal gluino mass of
100 GeV.

\section{Conclusions and Perspectives}
We have presented a first detailed study of the fragmentation functions
of SUSY QCD in the leading logarithmic approximation and presented,
for the first time, results for the f.f.'s
of all the partons into protons within a radiatively generated model.
Our analysis here has been quite conservative and focused on an intermediate
energy region
where small-x effects, linked to resummation and Regge behaviour,
which might invalidate a simple leading logarithmic (in $Q^2$) evolution,
may still be under control. Within
the obvious approximations involved in our study, we can still envision
that some of the features present in our results may become more pronounced
once we increase the original energy scale at which fragmentation occurs.
However, some variants, compared to the ordinary QCD scenario, are noticed.

An obvious question is why should we consider inclusion of supersymmetry
in the fragmentation function in the first place. After all the
fragmentation functions are measured at the GeV scale and
current experiments requires their extrapolation to energy
scales which are still below the SUSY threshold. In this
respect our analysis was motivated from the recent exciting
results from Ultra High Energy Cosmic Rays observations, which
may indicate the existence of a new hadronic scale, of the
order of $10^{11}{\rm GeV}$, or so. If the UHECR results
are substantiated further in the forthcoming experiments,
as we elaborated in the introduction, it precisely necessitates
the type of analysis that we pursued in this paper. It opens
up the need to extrapolate the QCD parameters to the new scale,
and the inclusion of any new degree of freedom which is
crossed in the extrapolation. As we discussed here this
also opens up a plethora of new issues and challenges for
the QCD extrapolation and analysis.

Although these variations are small at the energy probed in our numerical
analysis,
other sources of enhancement could set in at higher scales.
First of all, at a higher fragmenting scale, we expect more remarked
differences
in the overall distributions of the total fragmenting
multiplicities (above the 2 $\%$ level observed in Fig.~12),
but we also expect that various (and less inclusive)
correlations (such as energy-energy correlations and event
shape variables) may affect more substantially the structure
of a supersymmetric final state with respect to standard QCD.
As we approach the GZK cutoff, these features may become much more pronounced,
although the result has clearly to do with the distributions
and coupling of the supersymmetric constituents in the metastable state that
undergoes decay.

The natural question to ask is then:
what are the distributions of supersymmetric partons in a decaying metastable
state
of very large mass prior to fragmentation? At such large scales,
so far from the 1 TeV scale preferred by many standard supersymmetric models,
the Renormalization Group strategy
can be easily embraced to its fullest extent with sizeable consequences
on the low energy end. On this and other related issues
we hope to return in more detail in the near future.

\vspace{1cm}
\centerline{\bf Acknowledgements}
We thank S. Sarkar and R. Toldra for discussions.
C.C. and A.F. thank respectively
the Theory Groups at Oxford and Lecce for hospitality.
The work of C.C. is supported in part by INFN
(iniziativa specifica BARI-21) and by MURST. The work of A.F. is supported
by PPARC.

\section{Appendix 1. Input Functions}
For convenience we have included below the list
of f.f.'s taken from Ref.~\cite{kkp} that we have used

\begin{itemize}
\item $D_u^{p/\bar p}(x,\mu^2)=2D_d^{p/\bar p}(x,\mu^2)$:
\begin{eqnarray}
N&=&0.40211-0.21633\bar s-0.07045\bar s^2+0.07831\bar s^3\nonumber\\
\alpha&=&-0.85973+0.13987\bar s-0.82412\bar s^2+0.43114\bar s^3\nonumber\\
\beta&=&2.80160+0.78923\bar s-0.05344\bar s^2+0.01460\bar s^3\nonumber\\
\gamma&=& 0.05198\bar s-0.04623\bar s^2
\end{eqnarray}
\item $D_s^{p/\bar p}(x,\mu^2)$:
\begin{eqnarray}
N&=&4.07885-2.97392\bar s-0.92973\bar s^2+1.23517\bar s^3\nonumber\\
\alpha&=&-0.09735+0.25834\bar s-1.52246\bar s^2+0.77060\bar s^3\nonumber\\
\beta&=&4.99191+1.14379\bar s-0.85320\bar s^2+0.45607\bar s^3\nonumber\\
\gamma&=&0.07174 \bar s-0.08321\bar s^2
\end{eqnarray}
\item $D_c^{p/\bar p}(x,\mu^2)$:
\begin{eqnarray}
N&=&0.11061-0.07726\bar s_c+0.05422\bar s_c^2-0.03364\bar s_c^3\nonumber\\
\alpha&=&-1.54340-0.20804\bar s_c+0.29038\bar s_c^2-0.23662\bar s_c^3
\nonumber\\
\beta&=&2.20681+0.62274\bar s_c+0.29713\bar s_c^2-0.21861\bar s_c^3\nonumber\\
\gamma&=&0.00831\bar s_c+0.00065\bar s_c^2
\end{eqnarray}
\item $D_b^{p/\bar p}(x,\mu^2)$:
\begin{eqnarray}
N&=&40.0971-123.531\bar s_b+128.666\bar s_b^2-29.1808\bar s_b^3\nonumber\\
\alpha&=&0.74249-1.29639\bar s_b-3.65003\bar s_b^2+3.05340\bar s_b^3
\nonumber\\
\beta&=&12.3729-1.04932\bar s_b+0.34662\bar s_b^2-1.34412\bar s_b^3\nonumber\\
\gamma&=&-0.04290\bar s_b-0.30359\bar s_b^2
\end{eqnarray}
\item $D_g^{p/\bar p}(x,\mu^2)$:
\begin{eqnarray}
N&=&0.73953-1.64519\bar s+1.01189\bar s^2-0.10175\bar s^3\nonumber\\
\alpha&=&-0.76986-3.58787\bar s+13.8025\bar s^2-13.8902\bar s^3\nonumber\\
\beta&=&7.69079-2.84470\bar s-0.36719\bar s^2-2.21825\bar s^3\nonumber\\
\gamma&=&1.26515\bar s-1.96117\bar s^2
\end{eqnarray}
\end{itemize}

\section{Appendix 2. The Weights of the N=1 Kernels}
We briefly recall the numerical strategy
employed in this analysis. A more detailed description will be given elsewhere.
We just mention that the
radiative generation of supersymmetric distributions requires
special accuracy since these scaling violations grow up very slowly.
We define $\bar{P}(x)\equiv x P(x)$ and $\bar{A}(x)\equiv x A(x)$.
We also define the convolution product

\beq
J(x)\equiv\int_x^1 \frac{dy}{y}\left(\frac{x}{y}\right)
P\left(\frac{x}{y}\right)\bar{A}(y). \
\eeq
The integration interval in $y$ at any fixed x-value is
partitioned in an array of
increasing points ordered from left to right
$\left(x_0,x_1,x_2,...,x_n,x_{n+1}\right)$
with $x_0\equiv x$ and $x_{n+1}\equiv 1$ being the upper edge of the
integration
region. One constructs a rescaled array
$\left(x,x/x_n,...,x/x_2,x/x_1, 1 \right)$. We define
$s_i\equiv x/x_i$, and $s_{n+1}=x < s_n < s_{n-1}<... s_1 < s_0=1$.
We get
\beq
J(x)=\sum_{i=0}^N\int_{x_i}^{x_{i+1}}\frac{dy}{y}
\left(\frac{x}{y}\right) P\left(\frac{x}{y}\right)\bar{A}(y)
\eeq
At this point we introduce the linear interpolation
\beq
\bar{A}(y)=\left( 1- \frac{y - x_i}{x_{i+1}- x_i}\right)\bar{A}(x_i) +
\frac{y - x_i}{x_{i+1}-x_i}\bar{A}(x_{i+1})
\label{inter}
\eeq
and perform the integration on each subinterval with a change of variable
$y->x/y$ and replace the integral $J(x)$ with
its discrete approximation $J_N(x)$
to get
\beqa
J_N(x) &=& \bar{A}(x_0)\frac{1}{1- s_1}\int_{s_1}^1 \frac{dy}{y}P(y)(y - s_1)
\nonumber \\
&+& \sum_{i=1}^{N}\bar{A}(x_i) \frac{s_i}{s_i - s_{i+1}}
\int_{s_{i+1}}^{s_i} \frac{dy}{y}P(y)(y - s_{i+1})\nonumber \\
& -& \sum_{i=1}^{N}\bar{A}(x_i) \frac{s_i}{s_{i-1} - s_{i}}
\int_{s_{i}}^{s_{i-1}} \frac{dy}{y}P(y)(y - s_{i-1}) \nonumber \\
\eeqa
with the condition $\bar{A}(x_{N+1})=0$.
Introducing the coefficients  $W(x,x)$ and $W(x_i,x)$, the integral
is cast in the form
\beq
J_N(x)=W(x,x) \bar{A}(x) + \sum_{i=1}^{n} W(x_i,x)\bar{A}(x_i)
\eeq
where
\beqa
W(x,x) &=& \frac{1}{1-s_1} \int_{s_1}^1 \frac{dy}{y}(y- s_1)P(y), \nonumber \\
W(x_i,x) &=& \frac{s_i}{s_i- s_{i+1}}
\int_{s_{i+1}}^{s_i} \frac{dy}{y}\left( y - s_{i+1}\right) P(y) \nonumber \\
& -& \frac{s_i}{s_{i-1} - s_i}\int_{s_i}^{s_{i-1}}\frac{dy}{y}\left(
y - s_{i-1}\right) P(y).\nonumber \\
\eeqa

We recall that
\beq
  \int_0^1 dx \frac{f(x)}{(1-x)_+}=\int_0^1 {dy}\frac{f(y)- f(1)}{1-y}
\eeq
and that
\beq
 \frac{1}{(1-x)_+}\otimes f(x)\equiv
\int_x^1\frac{dy}{y}\frac{\,\,y f(y) - x f(x)}{y-x} + f(x)\log(1-x)
\eeq
as can can be shown quite straightforwardly.

We also introduce the expressions

\beqa
In_0(x) & = &
 \frac{1}{1- s_1} \log(s_1) + \log(1- s_1) \nonumber \\
\nonumber \\
Jn_i(x) & = & \frac{1}{s_i - s_{i +1}}
\left[ \log\left(\frac{1 - s_{i+1}}{1 - s_i}\right)
+ s_{i+1} \log\left(\frac{1- s_i}{1 - s_{i+1}}
\frac{s_{i+1}}{s_i}\right)\right]
\nonumber \\
Jnt_i(x) & = & \frac{1}{s_{i-1}- s_i}\left[ \log\left(\frac{1 - s_i}{1 -
s_{i-1}}\right)
+ s_{i-1}\log\left( \frac{s_i}{s_{i-1}}\right) + s_{i-1}\left(
\frac{1 - s_{i-1}}{1 - s_i}\right)\right],   \,\,\,\,\,\ i=2,3,..N \nonumber \\
Jnt_1(x) &=& \frac{1}{1- s_1}\log s_1. \nonumber \\
\eeqa

Using the linear interpolation formula (\ref{inter})  we get the
relation

\beqa
\int_x^1\frac{dy}{y} \frac{ y A_n(y) - x A_n(x)}{y-x} &=&
- \log(1-x) A_n(x) + A_n(x) In_0(x)\nonumber \\
&&  + \sum_{i=1}^N A_n(x_i) \left( Jn_i(x) - Jnt_i(x)\right)
\eeqa

which has been used for a fast
and accurate numerical implementation of the recursion relations.

\newpage

\begin{figure}
\centerline{\includegraphics[angle=0,width=.7\textwidth]{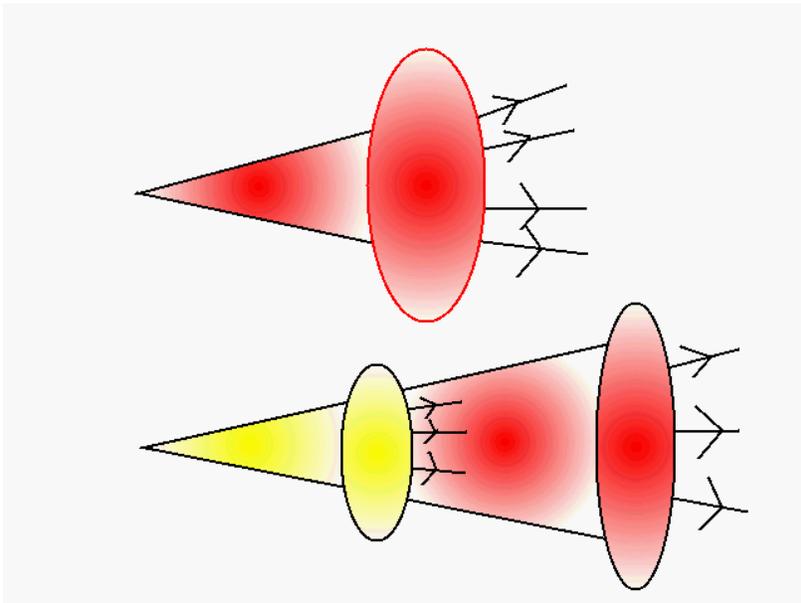}}
\caption{An illustration of the regular versus a mixed supersymmetric
evolution}
\end{figure}

\begin{figure}
\centerline{\includegraphics[angle=0,width=.8\textwidth]{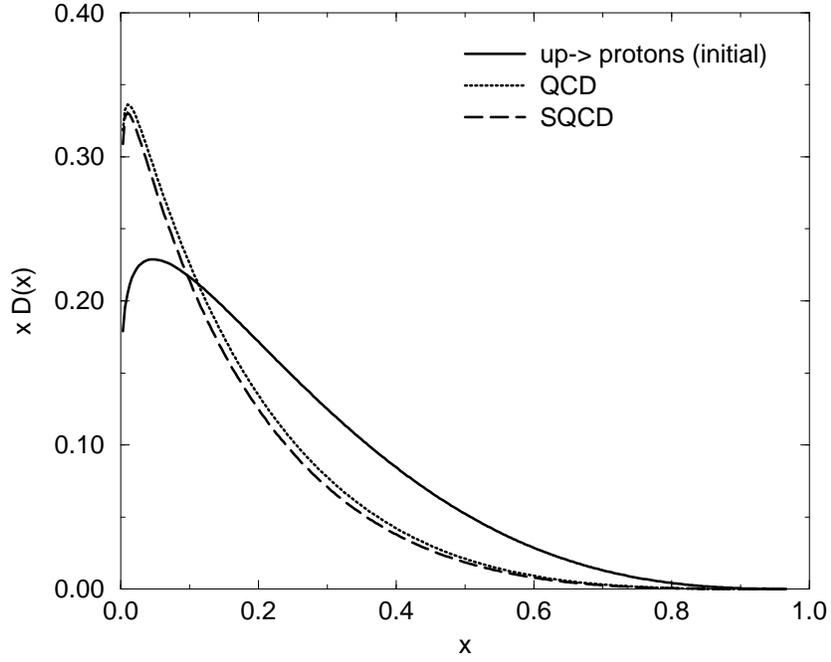}}
\centerline{\includegraphics[angle=0,width=.8\textwidth]{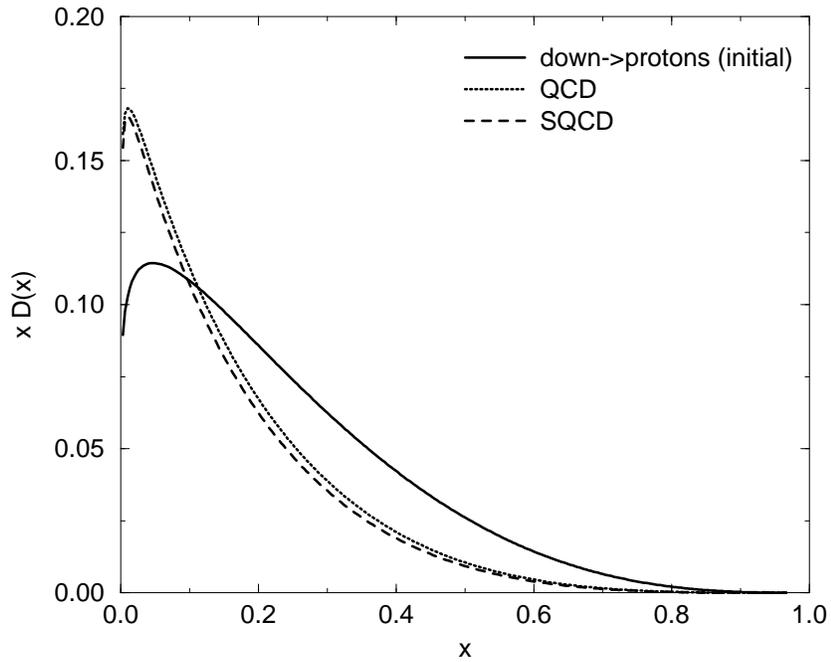}}
\caption{Fragmentation function for the
up (top graph) and down  quark (bottom graph)  $x D_{u,d}^{p,\bar{p}}(x,Q^2)$
at the lowest scale (input)
$Q_0=10$ GeV, and their QCD (or regular) and SQCD/QCD evolutions
with $Q_f=$$10^3$ GeV.The SUSY fragmentation scale is chosen to be $200$ GeV.
}.
\end{figure}

\begin{figure}
\centerline{\includegraphics[angle=0,width=.8\textwidth]{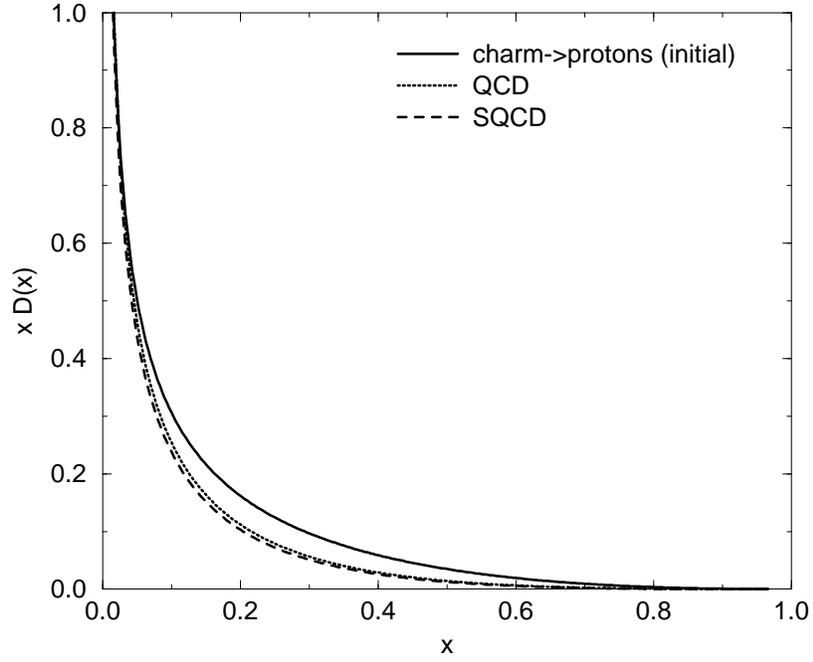}}
\centerline{\includegraphics[angle=0,width=.8\textwidth]{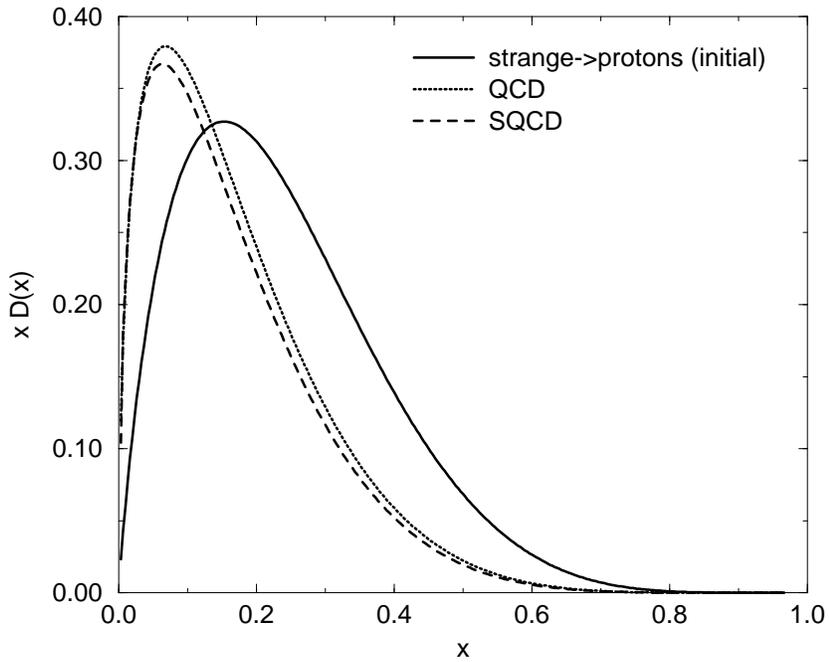}}
\caption{ Fragmentation functions into protons of
the charm and strange quarks $x D_{s,c}^{p,\bar{p}}(x,Q^2)$  at the lowest
scale (input)
$Q_0=10$ GeV, and its evolved QCD (regular) and SQCD/QCD evolutions
with $Q_f=$$10^3$ GeV.The SUSY fragmentation scale is chosen to be $200$ GeV.}
\end{figure}

\begin{figure}
\centerline{\includegraphics[angle=0,width=1.1\textwidth]{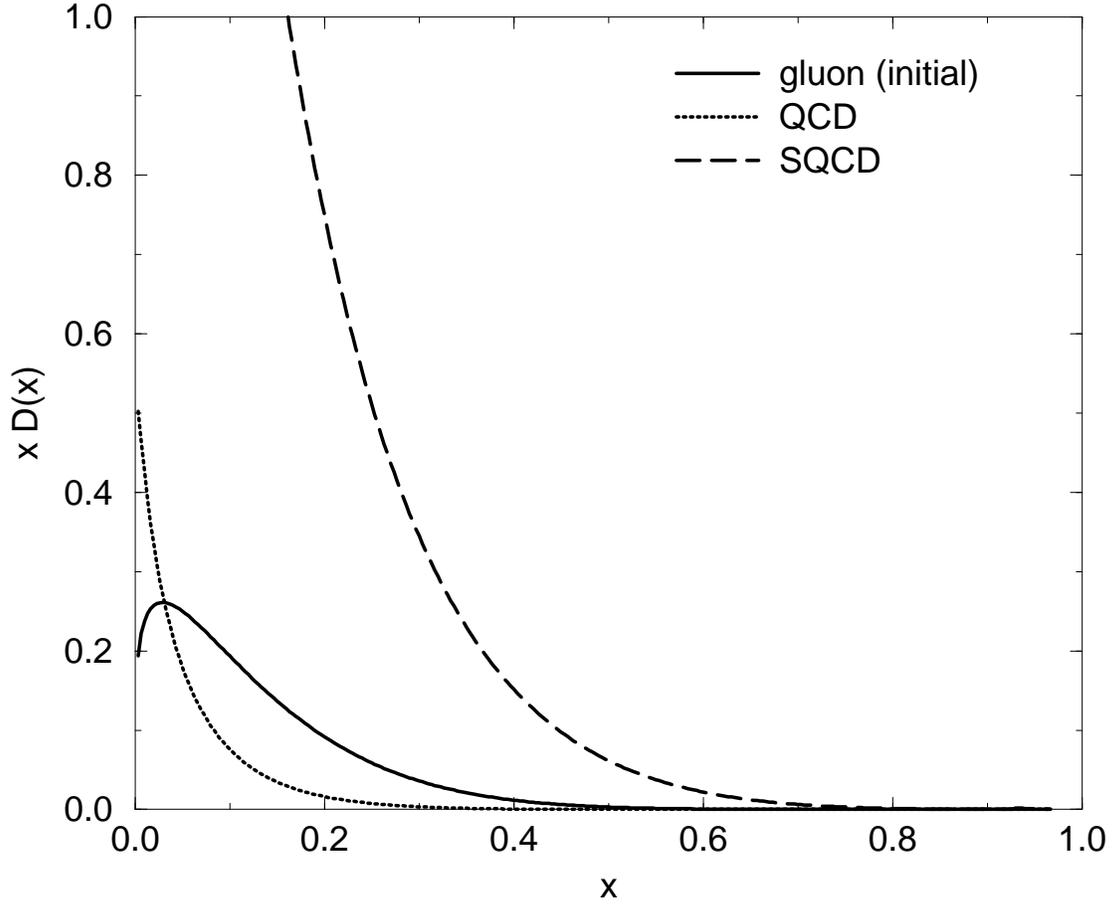}}
\caption{The gluon fragmentation function $x D_{g}^{p,\bar{p}}(x,Q^2)$  at the
lowest scale (input)
$Q_0=10$ GeV, and its evolved QCD (regular) and SQCD/QCD evolutions
with $Q_f=$$10^3$ GeV.The SUSY fragmentation scale is chosen to be $200$ GeV.}
\end{figure}
\newpage
\begin{figure}
\centerline{\includegraphics[angle=0,width=1.1\textwidth]{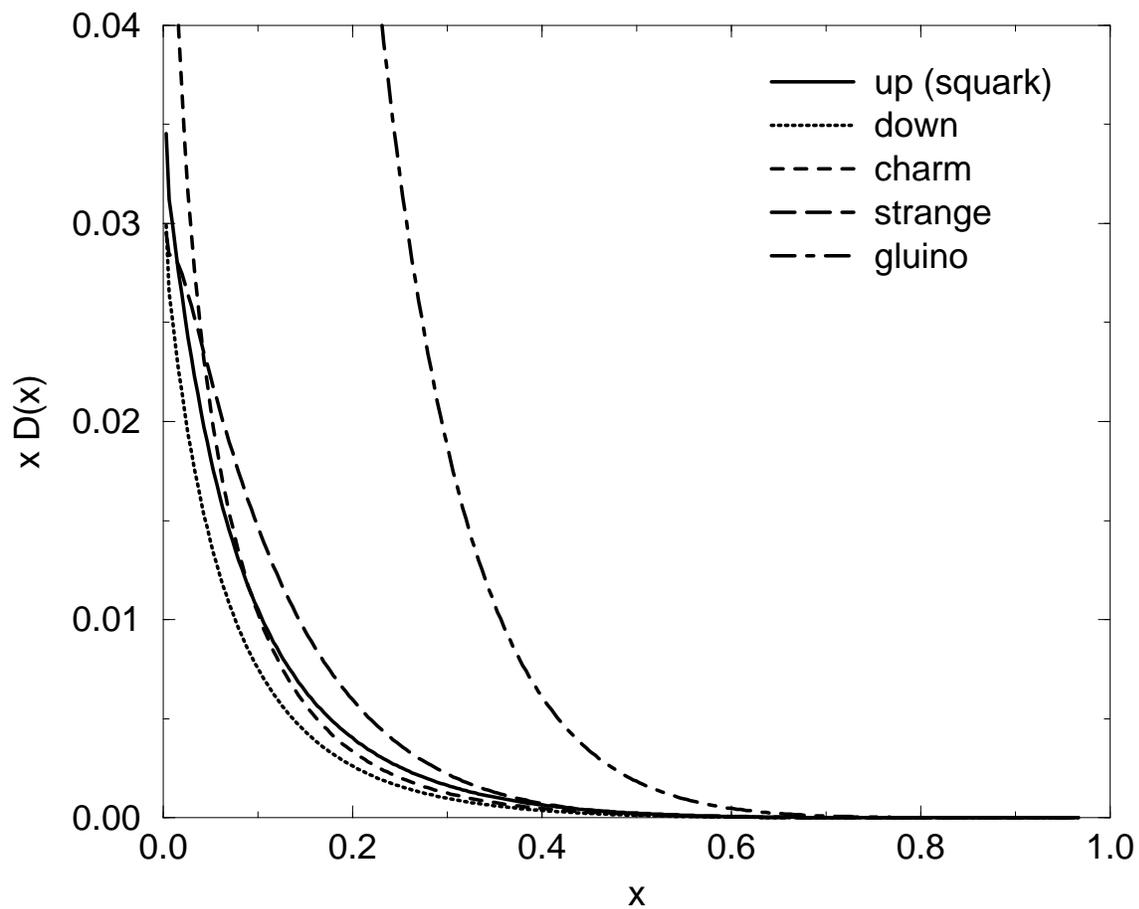}}
\caption{The fragmentation functions of squarks and gluino
 at the lowest scale (input)
$Q_0=10$ GeV, with $Q_f=10^3$ GeV and SUSY scale $200$ GeV.}
\end{figure}

\begin{figure}
\centerline{\includegraphics[angle=0,width=1.1\textwidth]{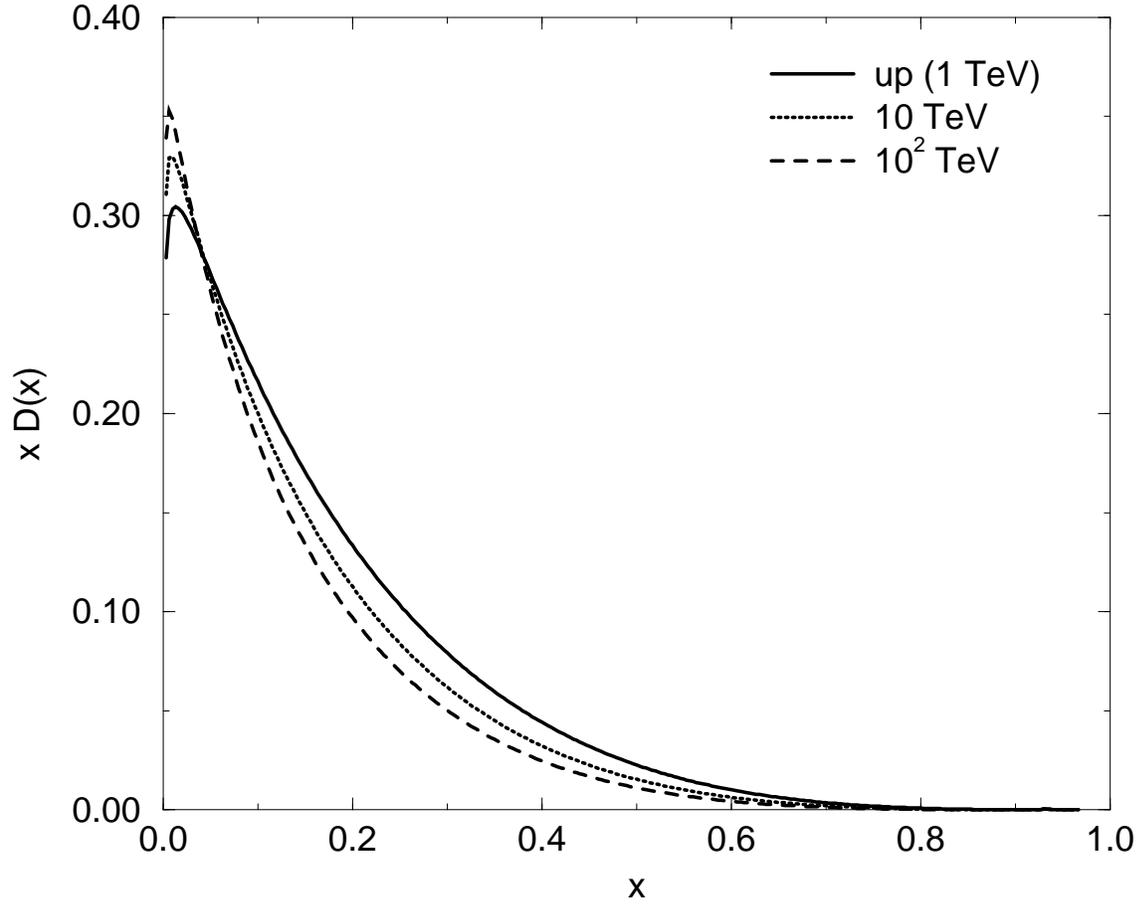}}
\caption{The up quark fragmentation function $x D_{u}^{p,\bar{p}}(x,Q^2)$
evolved with SQCD/QCD for varying final values of $Q_f$.
The SUSY fragmentation scale is chosen to be $200$ GeV.}
\end{figure}

\begin{figure}
\centerline{\includegraphics[angle=0,width=1.1\textwidth]{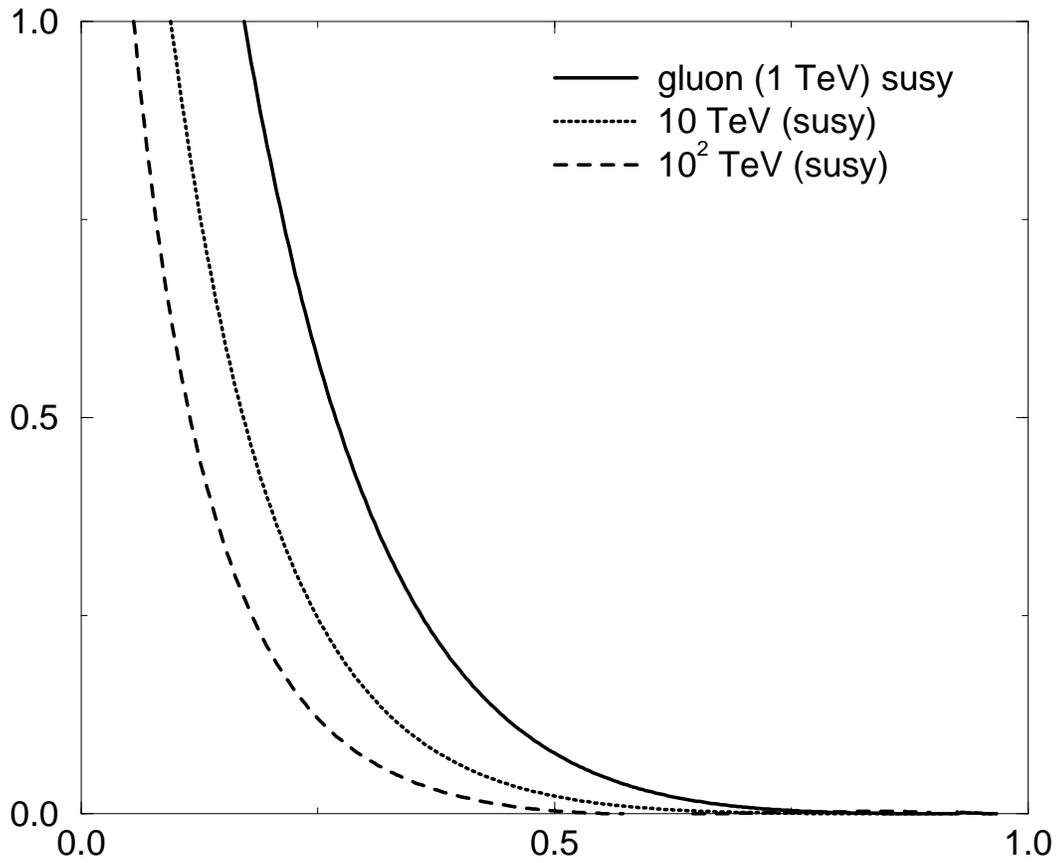}}
\caption{The gluon fragmentation function $x D_{g}^{p,\bar{p}}(x,Q^2)$
evolved regularly and according to SQCD/QCD for varying final values of $Q_f$.
The SUSY fragmentation scale is chosen to be $200$ GeV.}
\end{figure}

\begin{figure}
\centerline{\includegraphics[angle=0,width=1.1\textwidth]{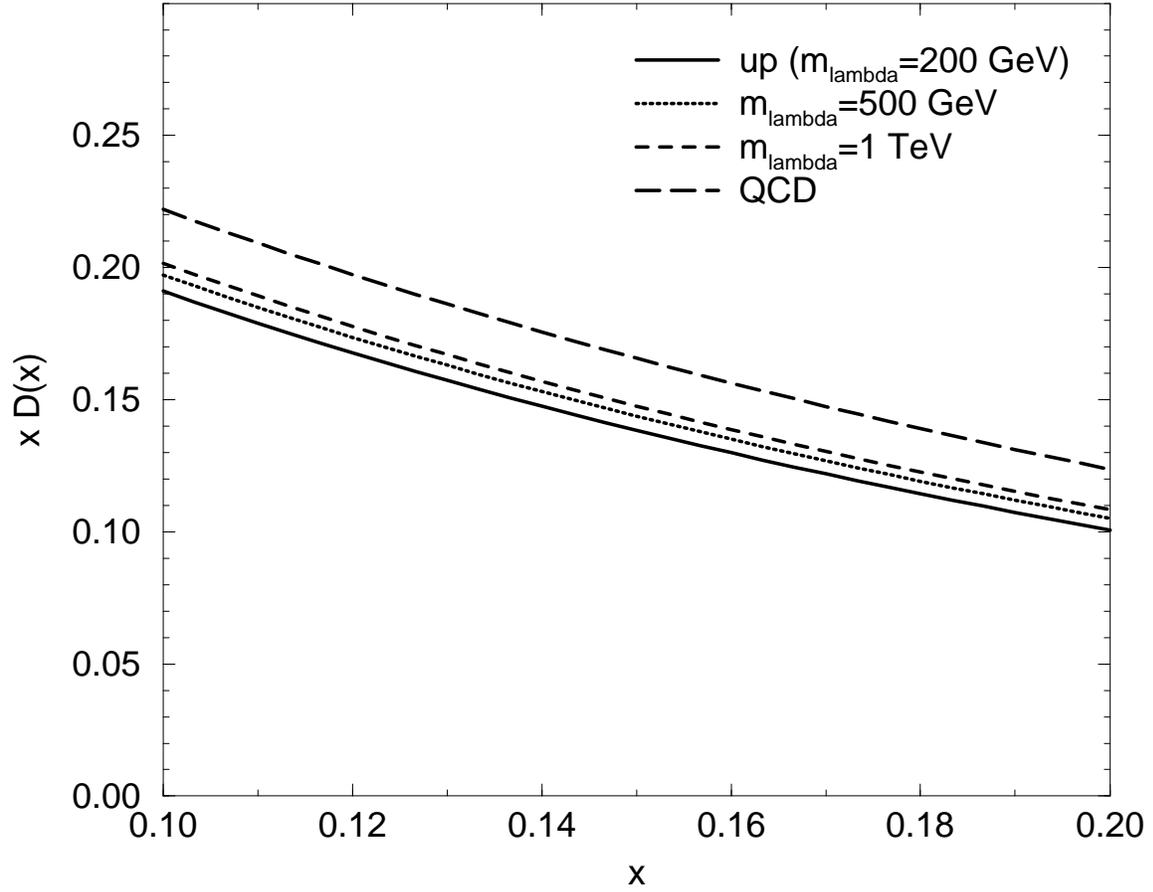}}
\caption{The up quark fragmentation function $x D_{u}^{p,\bar{p}}(x,Q^2)$
evolved regularly and according to SQCD/QCD for varying final values of the
SUSY scale $m_{2\lambda}$  and $Q_f=10^5$ GeV.
The SUSY fragmentation scale is chosen to be $200$ GeV.}
\end{figure}

\begin{figure}
\centerline{\includegraphics[angle=0,width=1.1\textwidth]{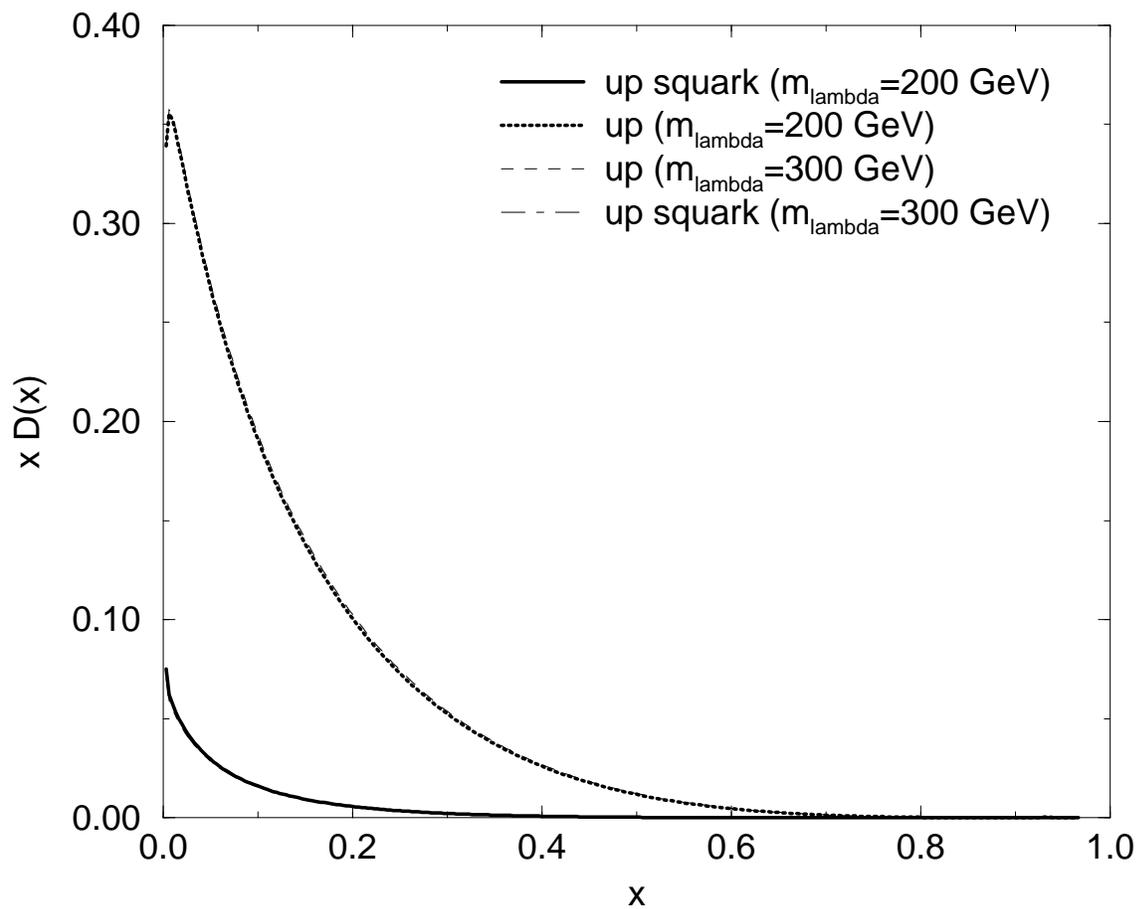}}
\caption{The quark and squark (up) fragmentation functions
for a varying $m_\lambda$ and a fixed large final scale $Q_f=10^5$ GeV.}
\end{figure}

\begin{figure}
\centerline{\includegraphics[angle=0,width=1.1\textwidth]{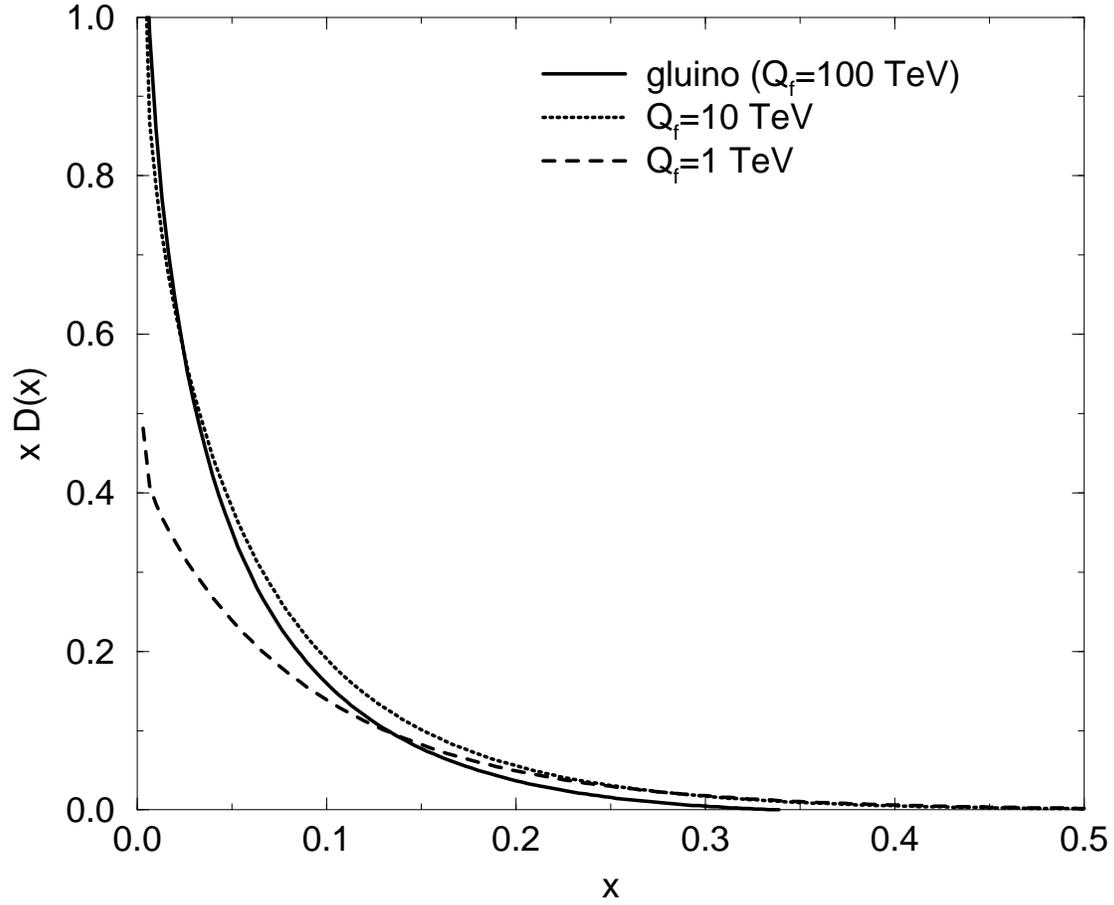}}
\caption{The gluino fragmentation function $x D_{u}^{p,\bar{p}}(x,Q^2)$
evolved in the region $200-Q_f$ GeV for different values of $Q_f$
The SUSY fragmentation scale is chosen to be $200$ GeV.}
\end{figure}

\begin{figure}
\centerline{\includegraphics[angle=0,width=1.1\textwidth]{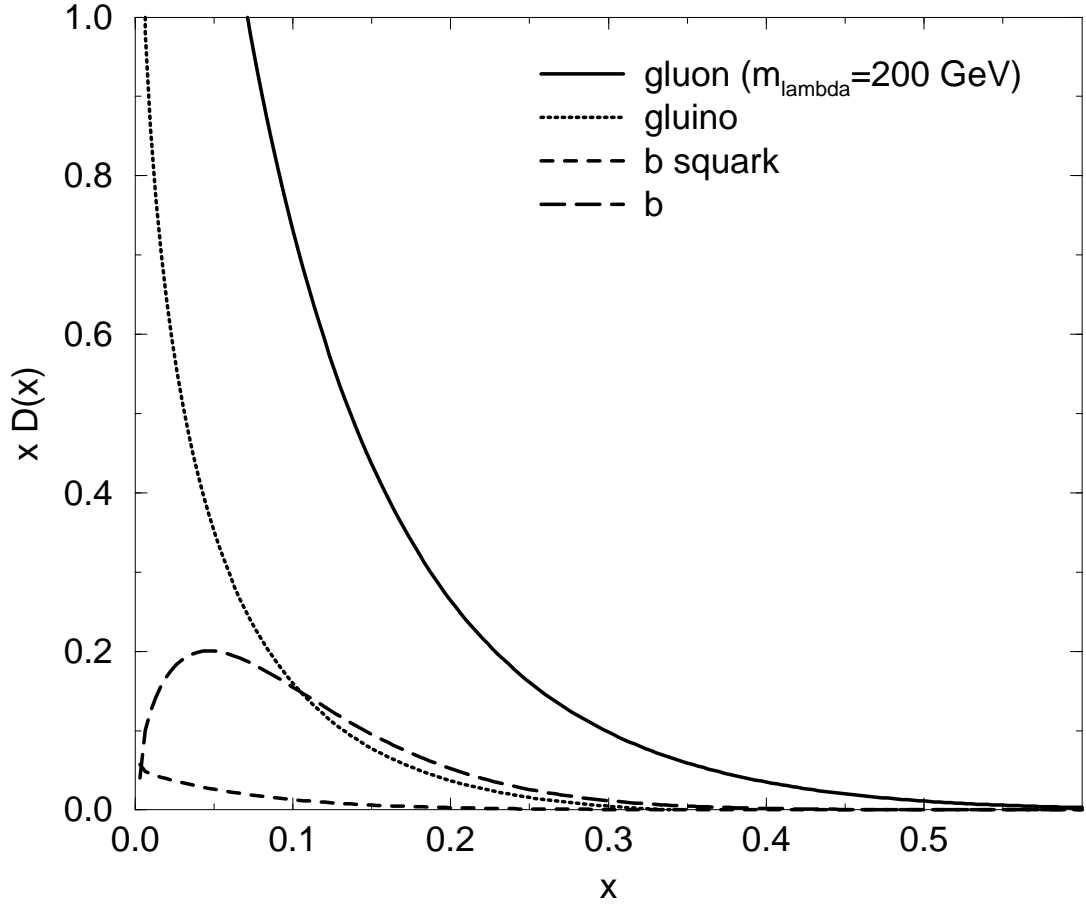}}
\caption{The quark and squark (bottom) fragmentation functions
together with those of gluon and gluino for $m_\lambda=100$ GeV and a fixed
large final scale $Q_f=10^5$ GeV.}
\end{figure}

\begin{figure}
\centerline{\includegraphics[angle=0,width=1.1\textwidth]{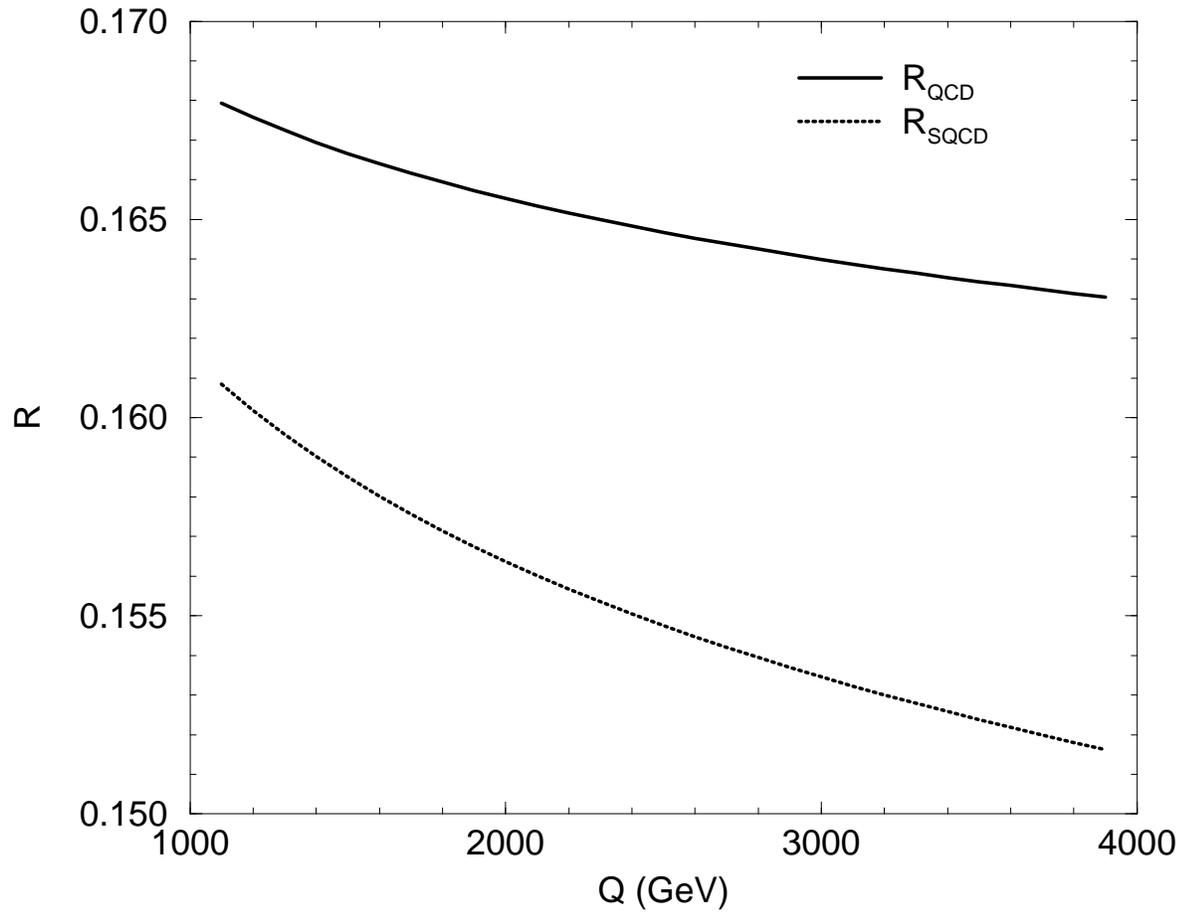}}
\caption{ $R_{QCD}$ and $R_{SQCD}$ for protons in the final state versus $Q$
(in $GeV$).}
\end{figure}

\end{document}